\documentclass[
aps,
prresearch,
reprint,
superscriptaddress,
amsmath,
amssymb,
longbibliography,
floatfix
]{revtex4-2}

\usepackage{bm}
\usepackage{mathtools}
\usepackage{microtype}
\usepackage{graphicx}

\newtheorem{lemma}{Lemma}
\newtheorem{proposition}{Proposition}

\usepackage{hyperref}
\hypersetup{hidelinks}

\begin{document}

\title{Operator uncertainty and cross-domain information in quantum signals:
Second-moment bounds for quantum waveform design}

\author{Mostafizur Rahaman Laskar}
\email{morl@kth.se}
\affiliation{Division of Information Science and Engineering, KTH Royal Institute of Technology, Stockholm, Sweden}
\affiliation{Quantum Information Physics Theory Research Team, RIKEN Center for Quantum Computing, Wako, Saitama 351-0198, Japan}

\author{Sajad Daei}
\affiliation{Division of Information Science and Engineering, KTH Royal Institute of Technology, Stockholm, Sweden}

\author{G\'abor Fodor}
\affiliation{Division of Information Science and Engineering, KTH Royal Institute of Technology, Stockholm, Sweden}
\affiliation{Ericsson, Sweden}

\author{Mikael Skoglund}
\affiliation{Division of Information Science and Engineering, KTH Royal Institute of Technology, Stockholm, Sweden}

\author{Franco Nori}
\affiliation{Quantum Information Physics Theory Research Team, RIKEN Center for Quantum Computing, Wako, Saitama 351-0198, Japan}
\affiliation{Department of Physics, University of Michigan,
Ann Arbor, Michigan 48109-1040, USA}

\begin{abstract}
We derive second-moment bounds for quantum waveform design by combining
entropic uncertainty relations with the covariance description of
continuous-variable systems. For any state with finite second moments and
entropies, the mutual information between the arrival time and the
transverse wave number, which are compatible observables from different
canonical pairs, is bounded above by the logarithm of twice the smaller of
the time-frequency and position-wave-number uncertainty products. We
demonstrate that a Gaussian wave packet with a bilinear space-time phase
saturates this bound for every admissible set of standard-deviation
budgets, and that the bound therefore provides a necessary and sufficient
condition for reaching a target information within the canonical model.
For this state, the joint measurement of the arrival time and the
transverse wave number is optimal among local Gaussian measurements,
yields the R\'enyi-2 entropy of the reduced state, and attains the quantum
Fisher information for the coupling at fixed input widths. We also show
that, with additive detector noise, timing jitter limits the recorded
information to a level set by the ratio of jitter to pulse duration. These
results provide a framework for designing spatiotemporally coupled quantum
signals, and we propose an optical realization based on pulse-front tilt
and a lens.
\end{abstract}

\maketitle

\section{Introduction}
\label{sec:intro}

Continuous-variable quantum signals encode information in temporal,
spectral, spatial, and spatial-frequency degrees of freedom. This
description connects photonic temporal modes \cite{Brecht2015,Fabre2022}
with the classical signal-domain picture, in which temporal waveforms map
to spectra and spatial fields map to spatial-frequency distributions
\cite{Gabor1946}. The temporal and spatial shape of a wave packet is
therefore a design variable in quantum communication, in particular for
optical quantum links based on temporal-mode encoding
\cite{Ansari2018,Raymer2020}. Multi-user continuous-variable networks use coherent-state quadratures and
coherent detection \cite{hajomer2024continuous}. Studies of pulse shaping have shown that temporal and spectral mode mismatch can degrade
the performance of the quantum channel
\cite{hajomer2024continuous,schlake2025pulse}. In the present work, the
continuous variables are the temporal and spatial coordinates of the wave
packet. The relations derived here constrain how temporal, spectral,
and spatial resources can be allocated before channel and network
constraints enter. Throughout, the term quantum signal refers to such a
quantum waveform. It should be distinguished from quantum signal processing in the sense
of polynomial transformations of block-encoded operators
\cite{low2017optimal,gilyen2019quantum,shendryk2025efficient} and from
signal-processing algorithms inspired by quantum measurement
\cite{eldar2002quantum}.

The temporal coordinate $T$ and its conjugate angular frequency $\nu$ form
one canonical pair, and the transverse position $X$ and transverse wave
number $K_x$ form another. Acting on the wave function $\psi(t,x)$ of the
wave packet, the coordinate operators $\hat T$ and $\hat X$ multiply by $t$
and $x$, respectively, and their conjugates are the derivatives
$\hat\nu=-i\partial_t$ and $\hat K_x=-i\partial_x$
(Sec.~\ref{sec:canonical_model}). These operators satisfy
\begin{equation*}
 [\hat T,\hat\nu]=i\openone,
 \qquad
 [\hat X,\hat K_x]=i\openone,
 \qquad
 [\hat T,\hat K_x]=0.
\end{equation*}
The first two commutators impose uncertainty relations, whereas the last
one permits a joint measurement of $(T,K_x)$. The uncertainty constraints
are instances of the Robertson and Robertson-Schr\"odinger relations
\cite{Robertson1929,Schrodinger1930}. Their phase-space treatment
follows the standard continuous-variable covariance formalism
\cite{Simon1994,braunstein2005quantum,Weedbrook2012}. We refer to the
Shannon mutual information between the outcomes of a compatible joint
measurement as the cross-domain information when the two observables belong
to different canonical pairs. This quantity is the standard mutual
information evaluated for a chosen pair of observables, rather than a new
information measure or a communication rate. It quantifies how much
one outcome reveals about the other. For instance, it bounds how well
a measurement of the transverse wave number can identify the arrival time of
the same wave packet. Although $T$ and $K_x$ commute, their outcomes can be
correlated through the state. Canonical uncertainty thus follows from
operator incompatibility, whereas the cross-domain information depends on
the prepared state and on the chosen pair of observables.

Entropic uncertainty relations with side information connect these two
notions \cite{BialynickiBirula1975,Coles2017}. In particular, the
information exclusion relation of Hall \cite{Hall1995,coles2014improved}
limits the information that a side variable can carry about two conjugate
observables. Position-momentum versions with quantum memory have been
discussed in Ref.~\cite{Furrer2014}. In a related direction, quantum
measurement theory separates intrinsic uncertainty from measurement error,
disturbance, and extracted information \cite{DresselNori2014}. Similarly,
quantum metrology separates generator fluctuations from the information
available for estimating a parameter. This information is quantified by
the classical and quantum Fisher information \cite{BraunsteinCaves1994,paris2009quantum,LeeNori2023}.
Time-frequency continuous variables provide a direct setting for quantum
metrology with single photons \cite{Donohue2018,Descamps2023}. In the
setting considered here, the side variable is a compatible observable of
the other canonical pair of the same wave packet. This raises three
questions: how canonical uncertainty constrains the information shared
by such observables, what maximum is attainable under prescribed
standard-deviation budgets, and which measurable observables attain it.

Here, we address these questions in a model that adds a transverse spatial
degree of freedom to the time-frequency description. This model allows the
above distinctions to be evaluated in one analytically tractable family. We
consider Gaussian states because their covariance matrices encode
canonical fluctuations and correlations between subsystems
\cite{Simon1994,Weedbrook2012,adesso2014continuous}. In addition, R\'enyi-2
quantities have covariance-based expressions and operational roles in
Gaussian information theory \cite{Adesso2012,MistaTatham2016,Lami2023}.
The coupling is the bilinear phase operator $\exp(i\chi\hat T\hat X)$,
which has the quadratic form of the continuous-variable controlled-$Z$
gate used for cluster states \cite{menicucci2006universal,gu2009quantum}.
In ultrafast optics, this phase is a first-order spatiotemporal coupling
that produces wave-front rotation \cite{Akturk2005,Guang2016}. Moreover,
spatiotemporal correlations of photon pairs \cite{gatti2009x} and
space-time wave packets \cite{kondakci2017diffraction,turo2026local} show
that the temporal and transverse degrees of freedom of light can be
coupled by design. Width ratios and Schmidt numbers of bipartite Gaussian
wave packets have been used as entanglement quantifiers
\cite{Fedorov2004,LawEberly2004}, and the Schmidt number of the
spatiotemporal spectrum has also been used to quantify the classical
non-separability of space-time wave packets \cite{Kondakci2019}. Within the
canonical model of Sec.~\ref{sec:canonical_model}, our analysis yields a
budget that specifies when a target correlation is feasible and provides a
Gaussian waveform that attains it.

\subsection{Contribution}

The main results of this work are as follows.

\begingroup
\setlength{\labelsep}{0.5em}
\settowidth{\leftmargini}{\textbullet}
\addtolength{\leftmargini}{\labelsep}
\begin{itemize}
\item We determine the maximum of the cross-domain information
$I(T;K_x)$ when the standard deviations of $T$, $\nu$, $X$, and $K_x$ are limited by prescribed upper
values, which we call budgets (Proposition~\ref{prop:optimal}). Over all
states with finite second moments and entropies (Sec.~\ref{sec:budget}),
this maximum is the logarithm of twice the smaller of the time-frequency
and position-wave-number budget products. A Gaussian wave packet with the
bilinear phase $\exp(i\chi\hat T\hat X)$ attains it for every set of
budgets allowed by the uncertainty relations, with $I(T;K_x)=\tfrac12\ln(1+g^2)$, where
$g=2\chi\sigma_t\sigma_x$ and $\sigma_t$ and $\sigma_x$ are the input
widths. The bound is therefore a necessary and sufficient condition for
reaching a target information.
\item For the same waveform, we show that the physical pair $(T,K_x)$
realizes the known optimum over local Gaussian measurements
\cite{mivsta2011measurement}. Its mutual information equals the R\'enyi-2
entropy of the reduced state and the logarithm of the Schmidt number, and
the same measurement attains the quantum Fisher information for the
coupling at fixed input widths.
\item We derive closed-form expressions for the recorded information and
Fisher information under finite detector resolution. These expressions
show that timing jitter \cite{gouzien2018quantum} limits the recorded
information to a saturation level, and they give feasibility conditions for
a target measured information. We also give a consistent estimator of the
coupling under calibrated jitter. In addition, measured second moments, including the beam-quality
factor, provide tomography-free upper limits on the information that a
spatial measurement can obtain about the arrival time.
\end{itemize}
\endgroup

Because the analysis concerns normalized wave packets and their Fourier
transforms, the same relations also hold for classical pulses with the
same spatiotemporal envelope \cite{Spreeuw1998,QianEberly2011}. The
results could therefore be tested with standard pulse-shaping optics.
Appendix~\ref{app:optical} outlines one possible realization based on
pulse-front tilt and a lens. For single photons, the same quantities could
be interpreted as entanglement between the temporal and spatial degrees of
freedom, provided that the photon statistics are verified independently.
Possible applications include pulse characterization, the estimation of
spatiotemporal couplings, and the assessment of spatial side information in
temporal-mode encodings (Sec.~\ref{sec:applications}).

\subsection{Relation to prior work}
\label{sec:prior}

Several ingredients of this work are known. Equation~\eqref{eq:exclX} is
the position and wave-number form of Hall's information exclusion relation
\cite{Hall1995}, with the side information supplied by the arrival time.
This relation follows from the Bia{\l}ynicki-Birula and Mycielski relation
\cite{BialynickiBirula1975}. When the position and the wave number are conditioned
on two different, incompatible observables of another system, the
corresponding inequality is the entropic steering criterion of Walborn
\emph{et al.} \cite{Walborn2011}, and entangled states can violate this
criterion. A
controlled-$Z$ interaction acting on a product of pure Gaussian states is
locally equivalent to a two-mode squeezed vacuum. For pure two-mode
squeezed states, local homodyne detection is the optimal local Gaussian
measurement for the classical mutual information
\cite{mivsta2011measurement}. The R\'enyi-2 entropy of a Gaussian state is
a function of its covariance matrix and is linked to the Shannon entropy of
its Wigner function \cite{Adesso2012}. The quantum Fisher information
of a pure-state family generated by an operator $\hat G$ is
$4\operatorname{Var}(\hat G)$ \cite{BraunsteinCaves1994}.

Building on these results, we take a compatible observable from the other
canonical pair as the side variable and bound $I(T;K_x)$ by both
uncertainty products. We also show that the correlations of $T$ with $X$, and of $K_x$
with $\nu$, are limited by the same uncertainty products. We obtain the
exact maximum under four standard-deviation budgets and show that it is
attained for every admissible budget. The bound therefore becomes a
feasibility condition. Furthermore, we identify a physically measurable
pair, the arrival time and the transverse wave number, that simultaneously
attains this bound, the optimum over local Gaussian measurements, the
R\'enyi-2 entropy, and the quantum Fisher information. We also give an
optical realization, detector-limited expressions, and tomography-free
bounds based on the beam-quality factor. To our knowledge, these results
have not been established together for the temporal and spatial degrees of
freedom of a wave packet.

Our analysis further connects three descriptions of a waveform: the
root-mean-square time-bandwidth and space-bandwidth products of signal
analysis \cite{Gabor1946}, the entropic uncertainty and exclusion
relations of quantum information
\cite{BialynickiBirula1975,Hall1995,Coles2017}, and the covariance-matrix
formalism of continuous-variable systems \cite{Simon1994,Weedbrook2012}.
Within this combined description, standard waveform operations act on the
vector
$\hat{\bm R}$ of Sec.~\ref{sec:canonical_model} either as displacements or
as symplectic maps. Time and frequency shifts are displacements and leave
the covariance matrix unchanged. Temporal chirps and dispersion are
quadratic phases in $t$ or $\nu$, and free paraxial propagation and thin
lenses are quadratic phases in $k_x$ or $x$. The bilinear shear of
Sec.~\ref{sec:shear} couples the two degrees of freedom. Each of these
operations maps
the covariance matrix as $V\mapsto SVS^{\top}$ with a symplectic $S$. The
budget of Eq.~\eqref{eq:resourcebound} can therefore be evaluated after
any sequence of such operations, using the second moments at the detection
plane. We carry out the analysis explicitly for the bilinear phase coupling
(Sec.~\ref{sec:shear}) and its optical realization by a lens
(Appendix~\ref{app:optical}). Other operations and their combinations are
left for future work.

\subsection{Organization of the paper}

The rest of this paper is organized as follows. In
Sec.~\ref{sec:canonical_model}, we introduce the canonical model and
specify the meaning of the arrival time. In Sec.~\ref{sec:budget}, we
derive the second-moment bound for arbitrary states and formulate the
standard-deviation budgets. In Sec.~\ref{sec:shear}, we introduce the
bilinear shear, and in Sec.~\ref{sec:information}, we show that the
sheared Gaussian state attains the bound and collect its
information-theoretic and metrological properties. In
Sec.~\ref{sec:detector}, we treat finite detector resolution and finite
samples. In Sec.~\ref{sec:discussion}, we discuss design rules, classical
and single-photon implementations, applications, and limitations. Finally,
we summarize our results in Sec.~\ref{sec:conclusion}.
Appendix~\ref{app:model} states the model assumptions for optical pulses,
and Appendix~\ref{app:optical} gives the optical realization.
Appendices~\ref{app:gaussian}, \ref{app:noisy}, and \ref{app:gaussianmeas}
contain derivations and proofs. Appendix~\ref{app:nongaussian} treats
examples beyond the Gaussian family, and Appendix~\ref{app:gate} describes
the discretized model and the numerical checks.

\section{Canonical Signal-Domain Model}
\label{sec:canonical_model}

Throughout this article, $\hat A$ denotes the operator associated with the
observable $A$, $\openone$ is the identity operator, and $\openone_n$ is
the $n\times n$ identity matrix. For example, $\hat T$ is the time
operator, $T$ is the arrival time obtained by measuring it, and $t$ is a
particular realization (value) of $T$; the same convention applies to $X$,
$K_x$, and $\nu$. For a density operator $\hat\rho$, the expectation value,
the fluctuation operator, the variance, and the symmetrized covariance are
\begin{align*}
 \langle\hat A\rangle_\rho&\coloneqq\mathrm{Tr}(\hat\rho\hat A),
 &
 \Delta\hat A&\coloneqq\hat A-\langle\hat A\rangle_\rho\openone,
 \\
 (\Delta A)^2&\coloneqq\langle(\Delta\hat A)^2\rangle_\rho,
 &
 C_{AB}&\coloneqq\tfrac12\langle\{\Delta\hat A,\Delta\hat B\}\rangle_\rho,
\end{align*}
where $\Delta A\ge0$ denotes the standard deviation and
$[\hat A,\hat B]\coloneqq\hat A\hat B-\hat B\hat A$ and
$\{\hat A,\hat B\}\coloneqq\hat A\hat B+\hat B\hat A$ are the commutator
and the anticommutator. The symbol $\coloneqq$ denotes a definition,
$\otimes$ the tensor product, $\oplus$ the direct sum of matrices,
$\mathrm{diag}(\cdot)$ a diagonal matrix, $^{\top}$ the transpose,
$^{\dagger}$ the Hermitian adjoint, and $M\succeq0$ that the Hermitian
matrix $M$ is positive semidefinite. For random variables, $\mathbb E$ and
$\operatorname{Var}$ denote expectation and variance, $A\sim\mathcal N(m,v)$
means that $A$ is Gaussian distributed with mean $m$ and variance $v$, and
$\approx$ denotes a leading-order approximation. The subscripts $T$ and
$X$ on states, kets, and covariance matrices label the temporal and spatial
subsystems. Natural logarithms are used, so information quantities are
measured in nats. Table~\ref{tab:notation} summarizes the principal
symbols.

\subsection{Wave packet and canonical operators}

We describe a continuous-variable signal by a normalized wave packet
$\psi(t,x)$ in the temporal coordinate $t$ and one transverse spatial
coordinate $x$. The state space is
$\mathcal H\coloneqq L^2(\mathbb R_t)\otimes L^2(\mathbb R_x)$, where
$L^2(\mathbb R_t)$ is the space of square-integrable functions of $t$. For
light, $\psi$ is the wave function of a single photon or the normalized
complex envelope of a classical pulse. The operators below act on this
single-excitation wave function and are not field-quadrature operators. The
Fourier representation of the wave packet is
\begin{align}
 \psi(t,x)
 &=
 \frac{1}{2\pi}
 \int_{\mathbb R} d\nu
 \int_{\mathbb R} dk_x\,
 \widetilde{\psi}(\nu,k_x)
 e^{\,i\nu t+i k_x x},
 \label{eq:envelope}
\end{align}
where $\widetilde{\psi}(\nu,k_x)$ is the Fourier transform of $\psi(t,x)$,
$\nu$ is the angular frequency conjugate to $t$, and $k_x$ is the
transverse wave number conjugate to $x$. The variables $t,x,\nu,k_x$ take
values in $\mathbb R$. The canonical operators and their commutators are
\begin{align*}
 \hat T&\coloneqq t,
 &
 \hat\nu&\coloneqq-i\frac{\partial}{\partial t},
 &
 [\hat T,\hat\nu]&=i\openone,
 \\
 \hat X&\coloneqq x,
 &
 \hat K_x&\coloneqq-i\frac{\partial}{\partial x},
 &
 [\hat X,\hat K_x]&=i\openone,
\end{align*}
and the operators of different pairs commute,
\begin{align*}
 [\hat T,\hat X]
 =
 [\hat T,\hat K_x]
 =
 [\hat\nu,\hat X]
 =
 [\hat\nu,\hat K_x]
 =0.
\end{align*}
The envelope is not assumed to be time-limited, band-limited, or of finite
spatial extent. The sign convention $e^{+i\nu t}$ in
Eq.~\eqref{eq:envelope} fixes $\hat\nu=-i\partial_t$. For a field whose
positive-frequency part varies as $e^{-i\omega t}$, it identifies
$\nu=\omega_c-\omega$, where $\omega_c$ is the carrier angular frequency,
so that positive $\nu$ corresponds to frequencies below the carrier. The
opposite sign reverses the frequency displacement in Eq.~\eqref{eq:heis}
and the sign of $C_{X\nu}$, and it leaves all variances, entropies, mutual
information, and Fisher information unchanged.

\subsection{Physical meaning of the arrival time}

Throughout, $T$ denotes the time at which the wave packet is detected in
a fixed transverse plane, which we call the arrival time. In Glauber's
theory of photodetection, the probability density of a detection event at
time $t$ and transverse position $x$ in that plane is proportional to the
normally ordered intensity
$\langle\hat E^{(-)}(t,x)\hat E^{(+)}(t,x)\rangle$ \cite{Glauber1963}. For
a narrowband paraxial pulse and an ideal broadband detector, this density
reduces to $|\psi(t,x)|^2$, where $\psi$ is the slowly varying envelope, so
$|\psi(t,x)|^2$ is the joint probability density of the arrival time and
the transverse position; Appendix~\ref{app:model} states the conditions
under which this holds. In this model $\hat T$ is the coordinate operator
of the envelope, and we do not identify it with a time operator conjugate
to the field Hamiltonian; for the arrival time in quantum mechanics, see
Ref.~\cite{MugaLeavens2000}. Statements below that a measurement is optimal, or that it attains a
quantum limit, therefore refer to measurements described within this
canonical model, not to all detection schemes of the quantized
electromagnetic field.

\subsection{Covariance matrix and Gaussian states}

For the ordered vector
$\hat{\bm R}\coloneqq(\hat T,\hat\nu,\hat X,\hat K_x)^{\top}$, we write
$[\hat R_j,\hat R_k]=iJ_{jk}\openone$, where $J=J_1\oplus J_1$ and
$J_1=\bigl(\begin{smallmatrix}0&1\\-1&0\end{smallmatrix}\bigr)$. The
covariance matrix contains the symmetrized second moments
\begin{equation*}
 V_{jk}\coloneqq\tfrac12\langle\{\Delta\hat R_j,\Delta\hat R_k\}\rangle_\rho .
\end{equation*}
In terms of the covariance matrix, the Robertson-Schr\"odinger uncertainty
relation \cite{Simon1994,Weedbrook2012} reads
\begin{align}
 V+\frac{i}{2}J
 &\succeq0.
 \label{eq:matrixUP}
\end{align}

The two tensor factors describe the temporal and the transverse spatial
degrees of freedom of one wave packet, not two independent modes of the
electromagnetic field. A single-photon state whose wave function
$\psi(t,x)$ is Gaussian has a Gaussian Wigner function in the canonical
variables $(T,\nu,X,K_x)$, although it is non-Gaussian with respect to the
field quadratures. The covariance-matrix and symplectic formalism of
Gaussian states applies to these canonical variables because they obey the
same commutation relations; the time-frequency and transverse spatial
variables of single photons have been treated as continuous variables in
this way \cite{Fabre2022,Tasca2011}.

\begin{table}[!htbp]
\caption{Principal notation. Widths denoted by $\sigma$ are standard
deviations of the input wave packet.}
\label{tab:notation}
\begin{ruledtabular}
\begin{tabular}{c p{0.70\columnwidth}}
Symbol & Meaning \\
\hline
$T,\nu$ & arrival time and angular frequency relative to the carrier, with the sign convention $\nu=\omega_c-\omega$ of Sec.~\ref{sec:canonical_model} \\
$X,K_x$ & transverse position and transverse wave number \\
$J$, $V$ & symplectic form and covariance matrix of $(T,\nu,X,K_x)$ \\
$V_T$, $V_X$ & covariance matrices of the reduced temporal and spatial states \\
$\mu_T$, $\zeta$ & purity and symplectic eigenvalue of the reduced temporal state \\
$K$ & Schmidt number of a pure state, $1/\mu_T$ \\
$\sigma_t,\sigma_x$ & temporal and spatial standard deviations of the input \\
$\chi,g$ & phase-coupling coefficient and dimensionless coupling \\
$S_\chi$ & symplectic matrix of the shear \\
$\beta_T,\beta_\nu,\beta_X,\beta_K$ & upper limits (budgets) on $\Delta T$, $\Delta\nu$, $\Delta X$, $\Delta K_x$ \\
$P_\star$ & smaller budget product, $\min\{\beta_T\beta_\nu,\beta_X\beta_K\}$ \\
$H(\cdot),I(\cdot\,;\cdot)$ & differential entropy and mutual information \\
$S_2$, $S_{\mathrm{vN}}$ & R\'enyi-2 and von Neumann entropies \\
$F_Q$, $F_{T,K_x}$ & quantum and classical Fisher information for $\chi$ \\
$\tau,\kappa$ & standard deviations of the timing jitter and of the wave-number noise \\
$a,b$ & normalized variances of the timing jitter and of the wave-number noise \\
$M$ & number of recorded outcomes
\end{tabular}
\end{ruledtabular}
\end{table}

For a Gaussian state of $n$ canonical pairs, the covariance matrix
determines the purity,
\begin{align*}
 \mu
 &\coloneqq
 \mathrm{Tr}(\hat\rho^2)
 =
 \frac{1}{\sqrt{\det(2V)}}.
\end{align*}
In the normalization $[\hat Q_\ell,\hat P_\ell]=i\openone$, a pure Gaussian
state has all symplectic eigenvalues equal to $1/2$, so that
$\det V=(1/4)^n$. For a pure Gaussian state of the two canonical pairs
considered here, $\det V=1/16$.

\section{Second-Moment Bound on the Cross-Domain Information}
\label{sec:budget}

\subsection{Entropic quantities and assumptions}

We consider an arbitrary state $\hat\rho$ on $\mathcal H$, pure or mixed,
Gaussian or non-Gaussian. Since $\hat T$ commutes with $\hat X$ and with
$\hat K_x$, the joint probability density functions $p(t,x)$ and
$p(t,k_x)$ exist, and since $\hat K_x$ commutes with $\hat\nu$, so does
$p(\nu,k_x)$. For a random variable $A$ with probability density function
$p_A$, the differential entropy, the conditional differential entropy, and
the mutual information are \cite{Cover2006}
\begin{align*}
 H(A)&\coloneqq-\int p_A\ln p_A,
 \\
 H(A|B)&\coloneqq H(A,B)-H(B),
 \\
 I(A;B)&\coloneqq H(A)-H(A|B).
\end{align*}
We assume finite second moments and finite differential entropies for all
marginal and joint distributions used below; statements about arbitrary
states refer to this class.

\subsection{Lemma 1: Exclusion relation for compatible cross-domain observables}

\begin{lemma}
\label{lem:exclusion}
Let $\hat\rho$ be a state on $\mathcal H$, pure or mixed, for which the
probability density functions of $T$, $\nu$, $X$, and $K_x$ and the joint
densities used below have finite second moments and finite differential
entropies. Then
\begin{align}
 I(T;K_x)+I(T;X)
 &\le H(X)+H(K_x)-\ln(\pi e)
 \nonumber\\
 &\le\ln(2\Delta X\Delta K_x),
 \label{eq:exclX}\\
 I(T;K_x)+I(\nu;K_x)
 &\le H(T)+H(\nu)-\ln(\pi e)
 \nonumber\\
 &\le\ln(2\Delta T\Delta\nu).
 \label{eq:exclT}
\end{align}
\end{lemma}

\emph{Proof.} For almost every outcome $t$ of $\hat T$, the conditional
state $\hat\rho_{X|t}\coloneqq\langle t|\hat\rho|t\rangle/p_T(t)$, where
$\langle t|\cdot|t\rangle$ acts on the temporal factor only, is a density
operator on $L^2(\mathbb R_x)$ whose position and wave-number probability
density functions are $p(x|t)$ and $p(k_x|t)$. The entropic uncertainty
relation of Bia{\l}ynicki-Birula and Mycielski \cite{BialynickiBirula1975}
for $[\hat X,\hat K_x]=i\openone$ gives
\begin{equation}
 H(X|T=t)+H(K_x|T=t)\ge\ln(\pi e),
 \label{eq:BBMcond}
\end{equation}
which holds for mixed states by concavity of the entropy \cite{Coles2017}.
Averaging Eq.~\eqref{eq:BBMcond} over $p_T$ yields
\begin{equation*}
 H(X|T)+H(K_x|T)\ge\ln(\pi e).
\end{equation*}
Inserting
\begin{align*}
 H(K_x|T)&=H(K_x)-I(T;K_x),
 \\
 H(X|T)&=H(X)-I(T;X)
\end{align*}
gives the first inequality in Eq.~\eqref{eq:exclX}. The second inequality
follows because the Gaussian probability density function maximizes the
differential entropy at fixed variance \cite{Cover2006},
\begin{equation}
 H(A)\le\tfrac12\ln\!\left[2\pi e\,(\Delta A)^2\right].
 \label{eq:maxent}
\end{equation}
Equation~\eqref{eq:exclT} follows in the same way after conditioning the
temporal degree of freedom on the outcome $k_x$ of $\hat K_x$.
\hfill$\square$

Equation~\eqref{eq:exclX} is Hall's information exclusion relation for
position and wave number \cite{Hall1995}, applied with the outcome of
$\hat T$ as the side variable, and Eq.~\eqref{eq:exclT} is the same
relation for the temporal pair with the outcome of $\hat K_x$ as the side
variable. Because each side variable is a single compatible observable,
the inequalities hold for every state in the class above. If $X$ and $K_x$
were instead conditioned on outcomes of two different, incompatible
observables of the temporal factor, the corresponding sum of conditional
entropies would be the entropic steering criterion of
Ref.~\cite{Walborn2011}, which entangled states can violate.

Since mutual information is nonnegative, every state satisfying these
assumptions obeys
\begin{align}
 I(T;K_x)
 &\le
 \min\left\{
 \ln(2\Delta T\Delta\nu),
 \ln(2\Delta X\Delta K_x)
 \right\}.
 \label{eq:generalbound}
\end{align}
In particular, if either uncertainty product takes its minimum value,
$\Delta T\Delta\nu=1/2$ or $\Delta X\Delta K_x=1/2$, the cross-domain
information vanishes. The available budget is the logarithm of the ratio
of an uncertainty product to its minimum value $1/2$, and any information
that $T$ shares with the position $X$ is deducted from the budget
available to $(T,K_x)$ in Eq.~\eqref{eq:exclX}. By the data-processing
inequality, the same bounds, with the standard deviations of the
unprocessed observables, hold when each of the two outcomes is processed
separately, including by additive noise and quantization. The second
moments are those at the detection plane at which $T$ and $K_x$ are
recorded; since $\Delta X$ changes under free propagation and $\Delta T$
under dispersion, the bound depends on this plane. Within the paraxial
model, free propagation leaves the joint distribution of $T$ and $K_x$
unchanged, so the bound $\ln(2\Delta X\Delta K_x)$ is tightest at the beam
waist (Sec.~\ref{sec:applications}).

\subsection{Standard-deviation budgets}

Let $\beta_T$, $\beta_\nu$, $\beta_X$, and $\beta_K$ be positive upper
limits on the four standard deviations, which we call budgets, and consider
all states with
\begin{align}
 \Delta T&\le\beta_T,
 &
 \Delta\nu&\le\beta_\nu,
 &
 \Delta X&\le\beta_X,
 &
 \Delta K_x&\le\beta_K.
 \label{eq:budgets}
\end{align}
The uncertainty relations require $\beta_T\beta_\nu\ge1/2$ and
$\beta_X\beta_K\ge1/2$ for this set to be nonempty.
Equation~\eqref{eq:generalbound} gives
\begin{align}
 I(T;K_x)
 &\le
 \min\left\{
 \ln(2\beta_T\beta_\nu),
 \ln(2\beta_X\beta_K)
 \right\}.
 \label{eq:resourcebound}
\end{align}
The budgets constrain standard deviations only;
Eq.~\eqref{eq:resourcebound} does not assume that the signal is
time-limited or band-limited, in time or in space. The products
$\Delta T\Delta\nu$ and $\Delta X\Delta K_x$ are the root-mean-square
time-bandwidth and space-bandwidth products of signal analysis
\cite{Gabor1946}, not the number of degrees of freedom of a time-limited
and band-limited signal. Section~\ref{sec:information} shows that
Eq.~\eqref{eq:resourcebound} is attained.

\section{Bilinear Temporal-Spatial Shear}
\label{sec:shear}

\subsection{Sheared Gaussian wave packet}

We consider a normalized wave packet whose envelope is separable in $t$ and
$x$, with the Gaussian envelope
\begin{align*}
 \psi_0(t,x)
 &\coloneqq
 \frac{1}{\sqrt{2\pi\sigma_t\sigma_x}}
 \exp\!\left(
 -\frac{t^2}{4\sigma_t^2}
 -\frac{x^2}{4\sigma_x^2}
 \right).
\end{align*}
Here $\sigma_t>0$ and $\sigma_x>0$ are the standard deviations of the
marginal probability density functions of $T$ and $X$, and the variances
of the four canonical variables are
\begin{align*}
 (\Delta T)^2&=\sigma_t^2,
 &
 (\Delta\nu)^2&=\frac{1}{4\sigma_t^2},
 \\
 (\Delta X)^2&=\sigma_x^2,
 &
 (\Delta K_x)^2&=\frac{1}{4\sigma_x^2}.
\end{align*}

We couple the temporal and spatial degrees of freedom by the bilinear phase
operator
\begin{align}
 \hat U_\chi
 &\coloneqq
 \exp(i\chi\hat T\hat X),
 \label{eq:unitary}
\end{align}
where $\chi\in\mathbb R$ is the phase-coupling coefficient, with units
$\mathrm{m}^{-1}\mathrm{s}^{-1}$ when $t$ is measured in seconds and $x$ in
meters. The operator $\hat U_\chi$ is unitary and has the form of the
continuous-variable controlled-$Z$ gate
\cite{menicucci2006universal,gu2009quantum}. Acting on the separable
input, it produces
\begin{align}
 \psi_\chi(t,x)
 &=
 \frac{1}{\sqrt{2\pi\sigma_t\sigma_x}}
 \exp\!\left[
 -\frac{t^2}{4\sigma_t^2}
 -\frac{x^2}{4\sigma_x^2}
 +i\chi tx
 \right].
 \label{eq:state}
\end{align}
The coupling leaves the modulus of the envelope unchanged and introduces
the phase $\chi tx$. For any nonzero $\chi$ the pure state $\psi_\chi$ is
not a product of a temporal and a spatial envelope. The conditional mean of
the transverse wave number is $\mathbb E[K_x|T=t]=\chi t$, so the local
wave-front tilt changes linearly across the pulse, which is the
wave-front rotation of ultrafast optics \cite{Akturk2005}.
Appendix~\ref{app:optical} shows that $\psi_\chi$ is produced at the back
focal plane of a lens from a pulse with pulse-front tilt.

\subsection{Heisenberg picture and symplectic matrix}

In the Heisenberg picture the operators transform as
\begin{align}
 \hat U_\chi^\dagger \hat T \hat U_\chi
 &=\hat T,
 &
 \hat U_\chi^\dagger \hat X \hat U_\chi
 &=\hat X,
 \nonumber\\
 \hat U_\chi^\dagger \hat\nu \hat U_\chi
 &=\hat\nu+\chi\hat X,
 &
 \hat U_\chi^\dagger \hat K_x \hat U_\chi
 &=\hat K_x+\chi\hat T.
 \label{eq:heis}
\end{align}
The angular frequency is displaced in proportion to the transverse
position, and the transverse wave number is displaced in proportion to the
temporal coordinate. In terms of $\hat{\bm R}$, Eq.~\eqref{eq:heis} reads
\begin{align}
 \hat{\bm R}'
 &=
 S_\chi\hat{\bm R},
 &
 S_\chi
 &\coloneqq
 \begin{pmatrix}
 1&0&0&0\\
 0&1&\chi&0\\
 0&0&1&0\\
 \chi&0&0&1
 \end{pmatrix}.
 \label{eq:S}
\end{align}
The matrix $S_\chi$ satisfies $S_\chi J S_\chi^{\top}=J$ and
$\det S_\chi=1$; it is a symplectic matrix describing a shear in phase
space.

\subsection{Second moments}

The covariance matrix of the input is
\begin{align*}
 V_0=\mathrm{diag}\bigl(\sigma_t^2,1/(4\sigma_t^2),
 \sigma_x^2,1/(4\sigma_x^2)\bigr),
\end{align*}
and that of the sheared state is
$V_\chi=S_\chi V_0S_\chi^{\top}$. The second moments become
\begin{align}
 (\Delta T)^2&=\sigma_t^2,
 &
 (\Delta X)^2&=\sigma_x^2,
 \nonumber\\
 (\Delta\nu)^2
 &=
 \frac{1}{4\sigma_t^2}+\chi^2\sigma_x^2,
 &
 (\Delta K_x)^2
 &=
 \frac{1}{4\sigma_x^2}+\chi^2\sigma_t^2,
 \nonumber\\
 C_{TK_x}&=\chi\sigma_t^2,
 &
 C_{X\nu}&=\chi\sigma_x^2,
 \label{eq:moments}
\end{align}
and all other covariances vanish. The coupling leaves the variances of $T$
and $X$ unchanged, increases the variances of their conjugate variables,
and generates the cross-domain covariances $C_{TK_x}$ and $C_{X\nu}$. Since
$\det S_\chi=1$,
\begin{align}
 \det V_\chi
 &=
 (\det S_\chi)^2\det V_0
 =
 \frac{1}{16},
 \label{eq:detV}
\end{align}
so the global state remains a pure Gaussian state.

We introduce the dimensionless coupling
\begin{align}
 g\coloneqq2\chi\sigma_t\sigma_x.
 \label{gvar}
\end{align}

\subsection{Standard form as a two-mode squeezed vacuum}

The relation to the standard form of a two-mode Gaussian covariance matrix
is explicit in the dimensionless quadratures
\begin{align*}
 Q_T&=\frac{T}{\sqrt{2}\sigma_t},
 &P_T&=\sqrt{2}\sigma_t\nu,
 \\
 Q_X&=\frac{X}{\sqrt{2}\sigma_x},
 &P_X&=\sqrt{2}\sigma_xK_x.
\end{align*}
The covariance matrix of the input is $\openone_4/2$, and the shear acts as
$P_T\mapsto P_T+gQ_X$ and $P_X\mapsto P_X+gQ_T$. Applying the local
squeezing transformation
$\mathrm{diag}\bigl((1+g^2)^{1/4},(1+g^2)^{-1/4}\bigr)$ to each degree of
freedom, followed by the phase-space rotation
$(Q_X,P_X)\mapsto(P_X,-Q_X)$ of the spatial degree of freedom, gives
\begin{align}
 V_{\mathrm{std}}
 =\frac12
 \begin{pmatrix}
 \gamma&0&g&0\\
 0&\gamma&0&-g\\
 g&0&\gamma&0\\
 0&-g&0&\gamma
 \end{pmatrix},
 \qquad
 \gamma\coloneqq\sqrt{1+g^2}.
 \label{eq:standardform}
\end{align}
Since $\gamma^2-g^2=1$, this is the covariance matrix of a two-mode
squeezed vacuum state in standard form, with squeezing parameter $r_s$
given by $\cosh(2r_s)=\gamma$ and $\sinh(2r_s)=g$
\cite{Simon1994,Weedbrook2012}. After the rotation, the rescaled $K_x$ is
the position quadrature of the spatial degree of freedom. The pair
$(T,K_x)$ therefore maps, up to local nonzero scale factors, to the two
correlated position quadratures of the standard form, and mutual
information is invariant under such local rescalings.

\section{Cross-Domain Information of the Sheared State}
\label{sec:information}

\subsection{Joint probability density and mutual information}

\begin{figure*}[!ht]
    \centering
    \includegraphics[width=0.90\textwidth]{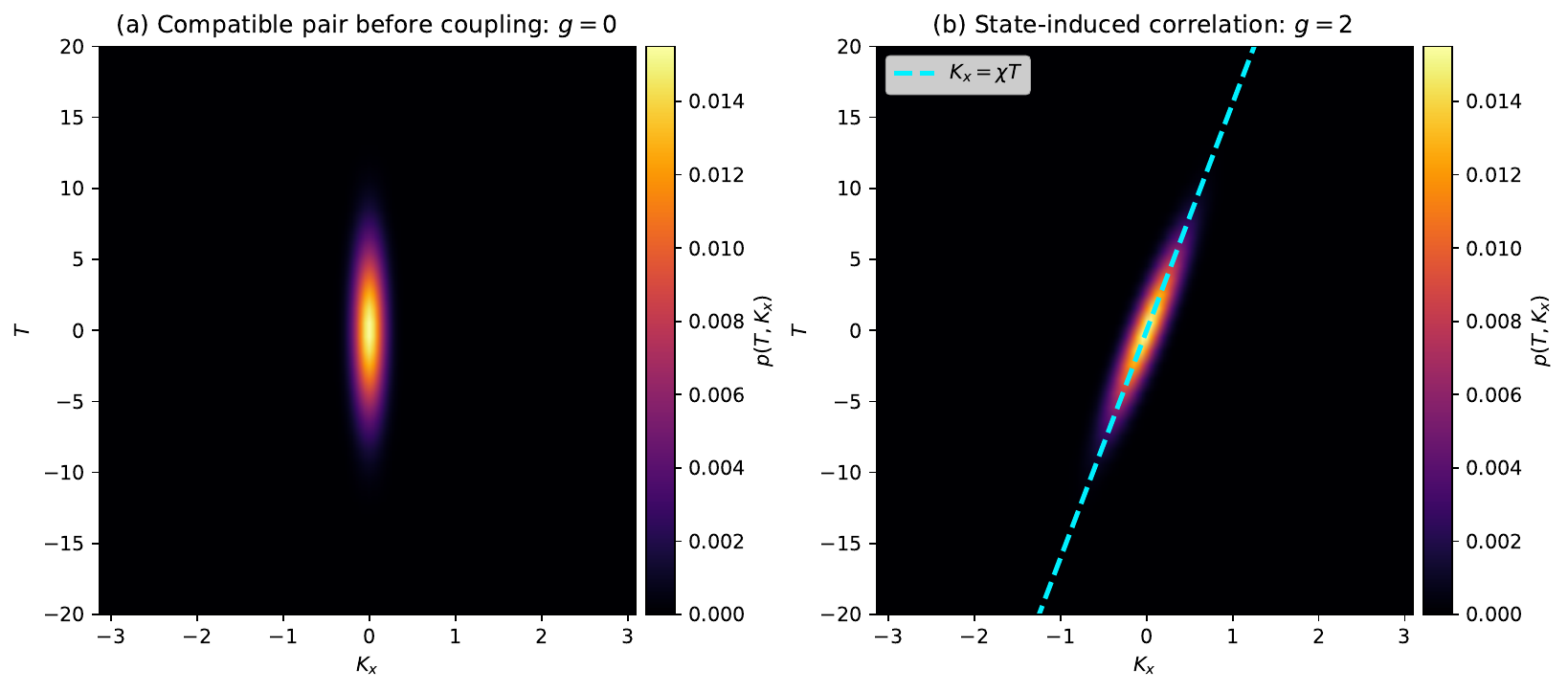}
    \caption{
    Joint probability mass function of the arrival time $T$ and the
    transverse wave number $K_x$ for the $128\times128$ discretized model
    of Appendix~\ref{app:gate}, with $d_t=d_x=1$ and
    $\sigma_t=\sigma_x=4$. The vertical axis is the temporal grid, the
    horizontal axis is the wave-number grid with spacing $2\pi/(Nd_x)$,
    and the color scale gives the probability of one grid cell.
    (a) For $g=0$, the outcomes of the compatible observables $T$ and $K_x$
    are statistically independent.
    (b) For $g=2$ ($\chi=1/16$), the bilinear phase produces a
    cross-domain correlation along the line $K_x=\chi T$. The commutator
    $[\hat T,\hat K_x]=0$ is unchanged; the statistical dependence is
    produced by the phase coupling, while the global covariance
    determinant keeps its pure-state value.
    }
    \label{fig:qiskit_joint}
\end{figure*}

A Fourier transform of Eq.~\eqref{eq:state} with respect to $x$ gives the
joint probability density function of the compatible pair,
\begin{align}
 p(t,k_x)
 &=
 \frac{e^{-t^2/(2\sigma_t^2)}}{\sqrt{2\pi}\sigma_t}
 \sqrt{\frac{2}{\pi}}\sigma_x
 \exp[-2\sigma_x^2(k_x-\chi t)^2].
 \label{eq:joint}
\end{align}
This is the linear Gaussian model
\begin{align}
 K_x&=\chi T+Y,
 &
 T&\sim\mathcal N(0,\sigma_t^2),
 &
 Y&\sim\mathcal N\!\left(0,\frac{1}{4\sigma_x^2}\right),
 \label{eq:linearlaw}
\end{align}
with $Y$ independent of $T$. The variable $Y$ is the transverse
wave-number spread of the input, fixed by $\sigma_x$; it is a property of
the state, not channel or receiver noise. The parameter $g^2$ is the ratio
of the variance of the conditional mean to the conditional variance,
\begin{align*}
 g^2
 &=
 \frac{\operatorname{Var}(\chi T)}{\operatorname{Var}(Y)}
 =4\chi^2\sigma_t^2\sigma_x^2,
\end{align*}
and plays the role of a signal-to-noise ratio (SNR). The Pearson
correlation coefficient is
\begin{align*}
 r_{T,K_x}
 =
 \frac{C_{TK_x}}{\Delta T\,\Delta K_x}
 =
 \frac{g}{\sqrt{1+g^2}},
\end{align*}
and for a bivariate Gaussian pair
$I(T;K_x)=-\tfrac12\ln(1-r_{T,K_x}^2)$ \cite{Cover2006}, which gives
\begin{align}
 I(T;K_x)
 &=
 \frac12\ln(1+g^2).
 \label{eq:MI}
\end{align}
The mutual information increases with $|g|$ while the determinant of the
global covariance matrix in Eq.~\eqref{eq:detV} stays at its pure-state
value. Figure~\ref{fig:qiskit_joint} shows the joint probability
distribution for $g=0$ and $g=2$, evaluated on the finite grid of
Appendix~\ref{app:gate}.

Equation~\eqref{eq:MI} has the form of the mutual information of a scalar
additive Gaussian noise channel with a Gaussian input and SNR $=g^2$. Here,
however, it is the information that one coordinate of the wave packet
carries about another coordinate of the same wave packet, and not a
channel capacity, since no message is encoded. For source coding the
correlation is a redundancy, since $H(T,K_x)=H(T)+H(K_x)-I(T;K_x)$. A
receiver that resolves only the transverse wave number obtains $I(T;K_x)$
nats of information about the arrival time. With
$\Delta(T|K_x)\coloneqq\sigma_t/\sqrt{1+g^2}$ the conditional standard
deviation of $T$ given $K_x$, the quantity
\begin{align}
 e^{I(T;K_x)}
 &=\sqrt{1+g^2}
 =\frac{\Delta T}{\Delta(T|K_x)}
 \label{eq:slots}
\end{align}
is the ratio of the prior to the posterior standard deviation of the
arrival time. It measures the effective improvement in temporal resolution
for this Gaussian model; counting distinguishable encoded time bins would
additionally require a binning and decision rule.

\subsection{Saturation of the second-moment bound}

Equation~\eqref{eq:moments} gives
\begin{align}
 1+g^2
 &=
 4(\Delta T)^2(\Delta\nu)^2
 =
 4(\Delta X)^2(\Delta K_x)^2,
 \label{eq:resourceproducts}
\end{align}
and Eq.~\eqref{eq:MI} becomes
\begin{align}
 I(T;K_x)
 &=
 \ln(2\Delta T\Delta\nu)
 =
 \ln(2\Delta X\Delta K_x).
 \label{eq:resourceidentity}
\end{align}
The sheared Gaussian state thus satisfies Eq.~\eqref{eq:generalbound} with
equality for both canonical pairs, and every inequality in the proof of
Lemma~\ref{lem:exclusion} is tight. The outcomes of $T$ and $X$ are
independent because $|\psi_\chi|^2=|\psi_0|^2$, and the outcomes of $\nu$
and $K_x$ are independent because they are jointly Gaussian with vanishing
covariance, so $I(T;X)=I(\nu;K_x)=0$. The conditional states are pure
Gaussian minimum-uncertainty states, and all marginal probability density
functions are Gaussian.

\subsection{Proposition 1: Maximum cross-domain information under budgets}


\begin{proposition}
\label{prop:optimal}
Let $\beta_T\beta_\nu\ge1/2$ and $\beta_X\beta_K\ge1/2$, and let
$P_\star\coloneqq\min\{\beta_T\beta_\nu,\beta_X\beta_K\}$. Over all states
satisfying the budgets in Eq.~\eqref{eq:budgets} and the assumptions of
Lemma~\ref{lem:exclusion},
\begin{align}
 \max_{\hat\rho}\,I(T;K_x)=\ln(2P_\star).
 \label{eq:optimal}
\end{align}
The maximum is attained by the sheared Gaussian state $\psi_\chi$ of
Eq.~\eqref{eq:state} with standard deviations chosen in
\begin{align}
 \frac{P_\star}{\beta_\nu}\le\sigma_t\le\beta_T,
 \qquad
 \frac{P_\star}{\beta_K}\le\sigma_x\le\beta_X,
 \label{eq:widthchoice}
\end{align}
and phase-coupling coefficient
\begin{align}
 |\chi|=\frac{\sqrt{4P_\star^2-1}}{2\sigma_t\sigma_x}.
 \label{eq:chichoice}
\end{align}
\end{proposition}

\emph{Proof.} The upper bound is Eq.~\eqref{eq:resourcebound}. The
intervals in Eq.~\eqref{eq:widthchoice} are nonempty because
$P_\star\le\beta_T\beta_\nu$ and $P_\star\le\beta_X\beta_K$. With
Eq.~\eqref{eq:chichoice}, $g^2=4P_\star^2-1$, and
Eq.~\eqref{eq:resourceproducts} gives
$\Delta T\Delta\nu=\Delta X\Delta K_x=P_\star$. The lower limits in
Eq.~\eqref{eq:widthchoice} ensure $\Delta\nu=P_\star/\sigma_t\le\beta_\nu$
and $\Delta K_x=P_\star/\sigma_x\le\beta_K$, so the state satisfies the
budgets, and Eq.~\eqref{eq:resourceidentity} gives
$I(T;K_x)=\ln(2P_\star)$. \hfill$\square$

Under the assumptions of Lemma~\ref{lem:exclusion}, no non-Gaussian or
mixed state exceeds the cross-domain information of the sheared Gaussian
state under the same budgets, and the smaller of the two budget products
sets the maximum. Because the sheared state has
$\Delta T\Delta\nu=\Delta X\Delta K_x$, only the smaller budget product is
binding. The lower limits of the width intervals in
Eq.~\eqref{eq:widthchoice} are the widths for which the conjugate widths
\begin{equation*}
 \Delta\nu=\frac{\sqrt{1+g^2}}{2\sigma_t},
 \qquad
 \Delta K_x=\frac{\sqrt{1+g^2}}{2\sigma_x}
\end{equation*}
remain within budget at $g^2=4P_\star^2-1$; for a target below the
maximum, the admissible intervals are wider (Sec.~\ref{sec:discussion}).
A fixed phase-coupling coefficient, or additional constraints from the
physical realization and the detector, further restrict the admissible
states, and Eq.~\eqref{eq:resourcebound} can then hold with strict
inequality. The sheared Gaussian state need not be the unique
maximizer.

\subsubsection{Role of the budget formulation}

The correlation that produces $I(T;K_x)$ is the same correlation that
enlarges $\Delta K_x$ and $\Delta\nu$ in Eq.~\eqref{eq:moments}, so the
coupling increases the same uncertainty products that bound the
information. For the sheared state alone,
Eq.~\eqref{eq:resourceproducts} already relates $I(T;K_x)$ to the
uncertainty products, whereas Proposition~\ref{prop:optimal} extends the
statement to every admissible state. Lemma~\ref{lem:exclusion} also shows
that the budget is shared: information that $T$ carries about $X$, or
that $K_x$ carries about $\nu$, is subtracted from what $(T,K_x)$ can
share. This identifies the correlations that compete with $I(T;K_x)$ for
the same budget. The lens of Appendix~\ref{app:optical} moves the correlation
from $(T,X)$ to $(T,K_x)$, whereas the space-time wave packets discussed
there carry correlations between $\nu$ and $K_x$. In addition, the bound
converts measured second moments of an unknown state into a
tomography-free upper limit on its cross-domain correlations, the form
used in Sec.~\ref{sec:applications}.

\subsection{Reduced state and R\'enyi-2 entropy}

The reduced temporal state $\hat\rho_T\coloneqq\mathrm{Tr}_X\hat\rho$,
where $\mathrm{Tr}_X$ is the partial trace over the spatial degree of
freedom, has the covariance matrix
\begin{align*}
 V_T
 &=
 \begin{pmatrix}
 \sigma_t^2 & 0\\
 0 & \dfrac{1}{4\sigma_t^2}+\chi^2\sigma_x^2
 \end{pmatrix},
 \qquad
 \det V_T
 =
 \frac14(1+g^2),
\end{align*}
and purity
\begin{align*}
 \mu_T
 &=
 \mathrm{Tr}(\hat\rho_T^2)
 =
 \frac{1}{\sqrt{1+g^2}}.
\end{align*}
At $g=0$ the state is a product state of the two degrees of freedom and
$\mu_T=1$; the purity of the reduced state decreases as $|g|$ increases.
For the pure global state this reflects non-separability between the two
degrees of freedom. For a mixed global state, a mixed reduced state alone
would not certify entanglement.

With the R\'enyi entropy of order~2,
$S_2(\hat\rho_T)\coloneqq-\ln\mathrm{Tr}(\hat\rho_T^2)$, the mutual
information of the sheared Gaussian state satisfies
\begin{align}
 I(T;K_x)
 &=
 S_2(\hat\rho_T)
 =
 -\ln\mu_T.
 \label{eq:identity}
\end{align}
Covariance-based R\'enyi-2 measures of Gaussian correlations were
introduced in Ref.~\cite{Adesso2012}, and their relation to Gaussian
intrinsic entanglement and to key distillation from two-mode field states
is studied in Refs.~\cite{MistaTatham2016,Lami2023}.
Equation~\eqref{eq:identity} specializes the covariance relation to the
present coordinates. The next proposition identifies the measurement that
realizes it.

\subsection{Proposition 2: Optimal local Gaussian measurement}

\begin{proposition}
\label{prop:gaussianmeas}
Let the temporal and spatial degrees of freedom of the sheared state
$\psi_\chi$ be measured by arbitrary local Gaussian measurements, including
homodyne limits. The mutual information of the two outcomes is at most
$\tfrac12\ln(1+g^2)$, and the joint measurement of $(T,K_x)$ attains this
value.
\end{proposition}

The proof is given in Appendix~\ref{app:gaussianmeas}. It identifies
$(T,K_x)$ as a physical realization of the known correlated-quadrature
optimum \cite{mivsta2011measurement}, so no further optimization over local
Gaussian measurements is required to obtain $S_2(\hat\rho_T)$. The ratio
in Eq.~\eqref{eq:slots} equals $1/\mu_T$. Ratios of unconditional to
conditional widths were introduced as entanglement quantifiers for
bipartite wave packets in Ref.~\cite{Fedorov2004}, and Schmidt-number
analyses of Gaussian bipartite amplitudes are given in
Ref.~\cite{LawEberly2004}. Here the two subsystems are degrees of freedom
of a single wave packet.

\subsection{Schmidt number and von Neumann entropy}

For a pure bipartite state the Schmidt number is
$K\coloneqq1/\mathrm{Tr}(\hat\rho_T^2)$ \cite{LawEberly2004}. For the
sheared Gaussian state, Eqs.~\eqref{eq:slots} and \eqref{eq:identity} give
\begin{align}
 K=\frac{1}{\mu_T}=\sqrt{1+g^2}
 =\frac{\Delta T}{\Delta(T|K_x)}
 =e^{I(T;K_x)}.
 \label{eq:schmidt}
\end{align}
The width ratio of Ref.~\cite{Fedorov2004}, the Schmidt number, and the
exponential of the cross-domain mutual information therefore coincide for
this family, and, for a state known to belong to it, the measured pair
$(T,K_x)$ determines the Schmidt number from a single correlation
coefficient. For every pure Gaussian state of the two degrees of freedom,
$\det V_T=\det V_X$ and
\begin{equation}
 K=2\sqrt{\det V_T}
 \le2\min\{\Delta T\Delta\nu,\Delta X\Delta K_x\},
 \label{eq:schmidtbound}
\end{equation}
because $\det V_T\le(\Delta T)^2(\Delta\nu)^2$ and
$\det V_X\le(\Delta X)^2(\Delta K_x)^2$. Under the budgets of
Eq.~\eqref{eq:budgets}, the Schmidt number of any pure Gaussian wave
packet is thus at most $2P_\star$, with equality for the state of
Proposition~\ref{prop:optimal}. For pure non-Gaussian wave packets,
$S_2(\hat\rho_T)=\ln K$ remains true by definition, whereas $I(T;K_x)$ need
not equal this quantity and the Gaussian width-ratio identity need not
hold (Appendix~\ref{app:nongaussian}).

The reduced state has the symplectic eigenvalue
$\zeta=\sqrt{\det V_T}=\tfrac12\sqrt{1+g^2}$, and its von Neumann entropy
is
\begin{align}
 S_{\mathrm{vN}}(\hat\rho_T)
 ={}&\left(\zeta+\frac12\right)
 \ln\!\left(\zeta+\frac12\right)
 \nonumber\\
 &-\left(\zeta-\frac12\right)
 \ln\!\left(\zeta-\frac12\right),
 \label{eq:vonNeumann}
\end{align}
with $0\ln0$ understood by continuity. For $g\neq0$,
$S_{\mathrm{vN}}(\hat\rho_T)>S_2(\hat\rho_T)=I(T;K_x)$, so the equality in
Eq.~\eqref{eq:identity} is specific to the R\'enyi entropy of order~2. The
entropy of entanglement of the pure global state is $S_{\mathrm{vN}}$, and
no local Gaussian measurement reaches it. Appendix~\ref{app:gaussian}
derives the joint probability density function and the purity of the
reduced state.

\subsection{Complementary correlations}

After reordering the phase-space coordinates as $(T,K_x,\nu,X)$, the
covariance matrix $V_\chi$ is block diagonal, and the Gaussian Wigner
function factorizes into two bivariate Gaussian functions. The pair
$(\nu,X)$ has the same correlation coefficient and mutual information as
$(T,K_x)$,
\begin{align}
 r_{\nu,X}
 &=\frac{g}{\sqrt{1+g^2}},
 &
 I(\nu;X)
 &=I(T;K_x).
 \label{eq:complementaryMI}
\end{align}
The R\'enyi-2 mutual information between the temporal and spatial degrees
of freedom, defined as a difference of R\'enyi-2 entropies
\cite{Adesso2012}, is
\begin{align}
 \mathcal I^{\mathrm{ed}}_2(T{:}X)
 &\coloneqq
 S_2(\hat\rho_T)+S_2(\hat\rho_X)-S_2(\hat\rho)
 \nonumber\\
 &=2S_2(\hat\rho_T)
 =I(T;K_x)+I(\nu;X),
 \label{eq:renyisplit}
\end{align}
where the superscript ``ed'' indicates the entropy-difference definition.
For Gaussian states, $\mathcal I^{\mathrm{ed}}_2$ equals the Shannon mutual
information between the temporal and spatial arguments of the Wigner
function \cite{Adesso2012}. Here the Wigner function factorizes into the
$(T,K_x)$ and $(\nu,X)$ blocks, which is why it splits into the mutual
information of the pair $(T,K_x)$ and that of the pair $(\nu,X)$. The decomposition in
Eq.~\eqref{eq:renyisplit} is specific to the pure sheared Gaussian state.
The two pairs cannot be measured jointly, because $T$ and $\nu$ are
conjugate, as are $X$ and $K_x$.

The correlation increases both uncertainty products. From
Eq.~\eqref{eq:moments},
\begin{align}
 \Delta T\Delta\nu
 =\Delta X\Delta K_x
 &=\frac12\sqrt{1+g^2}
 =\frac{1}{2\mu_T}.
 \label{eq:localuncertainty}
\end{align}
The determinant of the global covariance matrix remains $1/16$, while the
uncertainty product of each canonical pair exceeds its minimum value $1/2$
when $g\neq0$.

\subsection{Fisher-optimal estimation of the coupling}

The same compatible measurement attains the quantum Fisher information for
estimating the phase-coupling coefficient at fixed $\sigma_t$ and
$\sigma_x$. Writing $\hat U_\chi=\exp(i\chi\hat G)$ with the generator
$\hat G\coloneqq\hat T\hat X$, the quantum Fisher information of the pure
state family $\hat U_\chi\psi_0$ is
\cite{BraunsteinCaves1994,paris2009quantum}
\begin{align}
 F_Q
 &=4\operatorname{Var}_{\psi_0}(\hat G)
 =4\sigma_t^2\sigma_x^2.
 \label{eq:qfi}
\end{align}
For the model in Eq.~\eqref{eq:linearlaw}, the marginal probability density
function of $T$ does not depend on $\chi$, and the classical Fisher
information is
\begin{align}
 F_{T,K_x}(\chi)
 &=
 \frac{\mathbb E[T^2]}{\operatorname{Var}(Y)}
 =4\sigma_t^2\sigma_x^2
 =F_Q.
 \label{eq:classicalFI}
\end{align}
The classical Fisher information of the fixed compatible measurement thus
equals the quantum Fisher information, so the corresponding Cram\'er-Rao
bound coincides with the quantum Cram\'er-Rao bound. The measurement does
not depend on $\chi$, and no adaptive scheme is required. By contrast, the
probability density function of $(T,X)$ is independent of $\chi$ and
carries no Fisher information. The bound is the quantum limit for this
separable probe within the canonical model of
Sec.~\ref{sec:canonical_model}. It involves no scaling advantage, and a
classical pulse gives the same Fisher information per detected photon.

For any centered separable probe, $\psi_0(t,x)=f(t)h(x)$ with
$\langle\hat T\rangle=\langle\hat X\rangle=0$, one has
\begin{equation*}
 \operatorname{Var}(\hat G)=(\Delta T)^2(\Delta X)^2 .
\end{equation*}
Since the shear leaves $\Delta T$ and $\Delta X$ unchanged, $F_Q$ is at
most $4\beta_T^2\beta_X^2$ under the budgets of Eq.~\eqref{eq:budgets}.
Probes with correlated temporal and spatial fluctuations can have
$\operatorname{Var}(\hat T\hat X)>\sigma_t^2\sigma_x^2$ and hence a larger
$F_Q$; an optimization over probe states is beyond the scope of this work. Since $g^2=\chi^2F_Q$,
Eq.~\eqref{eq:MI} can be written as
\begin{equation*}
 I(T;K_x)=\tfrac12\ln\!\left(1+\chi^2F_Q\right).
\end{equation*}

\subsection{Summary of the identities for the sheared state}

For the pure sheared Gaussian state, the relations of this section are
collected as
\begin{equation}
\boxed{
\begin{aligned}
 F_{T,K_x}(\chi)&=F_Q,\\
 I(T;K_x)&=S_2(\hat\rho_T)=\ln K\\
 &=\ln(2\Delta T\Delta\nu)\\
 &=\ln(2\Delta X\Delta K_x)\\
 &=\frac12\ln\!\left(1+\chi^2F_Q\right).
\end{aligned}
}
\label{eq:centralchain}
\end{equation}
The equalities with the two uncertainty products express the saturation of
Lemma~\ref{lem:exclusion}.

\subsection{Scope beyond the Gaussian family}

Lemma~\ref{lem:exclusion} and Eq.~\eqref{eq:resourcebound} hold for every
state in the class defined in Sec.~\ref{sec:budget}; the equalities in
Eq.~\eqref{eq:centralchain} do not. The mutual information of a
measurement is a property of the selected outcomes, whereas entanglement is
a property of the full state. A separable mixed state can have nonzero
mutual information through classical correlations, and an entangled state
can give zero mutual information for a particular compatible measurement;
thus, without additional assumptions on the state, $I(T;K_x)$ alone
neither certifies nor quantifies entanglement.
Appendix~\ref{app:nongaussian} gives a continuous-variable non-Gaussian
envelope for which $I(T;K_x)$ lies strictly below
Eq.~\eqref{eq:generalbound} and differs from $S_2(\hat\rho_T)$, and a
two-level example in which $I(T;K_x)$ equals $S_{\mathrm{vN}}$.

\section{Finite Detector Resolution and Finite Samples}
\label{sec:detector}

\subsection{Detector model}

A physical receiver resolves the arrival time and the transverse wave
number with finite resolution. We model the recorded outcomes as
\begin{align}
 T'&=T+\xi_t,
 &
 K_x'&=K_x+\xi_k,
 \label{eq:noisemodel}
\end{align}
where $\xi_t\sim\mathcal N(0,\tau^2)$ and $\xi_k\sim\mathcal N(0,\kappa^2)$
are independent of each other and of $(T,K_x)$. The standard deviation
$\tau$ represents the timing jitter \cite{gouzien2018quantum}, and $\kappa$
represents the wave-number resolution. We define the normalized noise
variances
\begin{align}
 a\coloneqq\frac{\tau^2}{\sigma_t^2},
 \qquad
 b\coloneqq4\sigma_x^2\kappa^2,
 \label{eq:ab}
\end{align}
so that $b$ is the variance of the wave-number noise in units of
$\operatorname{Var}(Y)$.

\subsection{Proposition 3: Information and Fisher information of recorded outcomes}

\begin{proposition}
\label{prop:noisy}
For the sheared Gaussian state and the detector model in
Eq.~\eqref{eq:noisemodel}, let $\Lambda\coloneqq(1+a)(1+b)+ag^2$. For the
Fisher information with respect to $\chi$, the input widths
$\sigma_t,\sigma_x$ and the detector-noise widths $\tau,\kappa$ are
treated as known and held fixed independently of $\chi$. The recorded
outcomes have
\begin{align}
 I(T';K_x')
 &=\frac12\ln\frac{(1+a)(1+b+g^2)}{\Lambda},
 \label{eq:noisyMI}\\
 F_{T',K_x'}(\chi)
 &=F_Q\,\frac{\Lambda+2a^2g^2}{\Lambda^2}.
 \label{eq:noisyFI}
\end{align}
\end{proposition}

\begin{figure*}[!ht]
    \centering
    \includegraphics[width=0.85\textwidth]{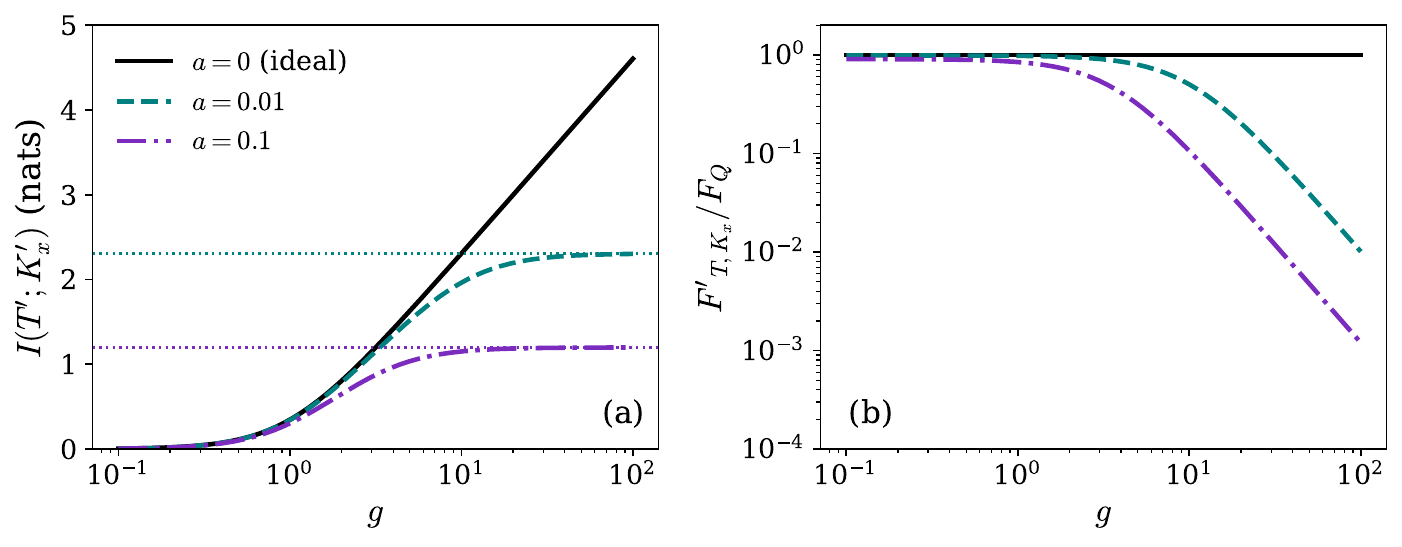}
    \caption{
    Measurement of the sheared Gaussian state with finite detector
    resolution, for normalized timing-jitter variance
    $a=\tau^2/\sigma_t^2$ and normalized wave-number-noise variance
    $b=4\sigma_x^2\kappa^2$. All curves use $b=0$ and the indicated
    values of $a$, isolating the effect of timing jitter.
    (a) Mutual information of the recorded outcomes,
    Eq.~\eqref{eq:noisyMI}. Without jitter it grows without limit; with
    jitter it approaches the level $\tfrac12\ln(1+1/a)$ of
    Eq.~\eqref{eq:saturation} (horizontal lines), which does not depend
    on $b$.
    (b) Classical Fisher information of the recorded outcomes relative to
    the quantum Fisher information, Eq.~\eqref{eq:noisyFI}. Jitter
    reduces the relative Fisher information at large coupling.
    }
    \label{fig:detector}
\end{figure*}

The derivation is given in Appendix~\ref{app:noisy}. For $a=b=0$ the
noiseless results in Eqs.~\eqref{eq:MI} and \eqref{eq:classicalFI} are
recovered. Figure~\ref{fig:detector} shows both quantities as functions of
$g$.

\subsection{Saturation by timing jitter}

Without timing jitter, $a=0$, the wave-number noise reduces the effective
signal-to-noise ratio from $g^2$ to $g^2/(1+b)$,
\begin{align*}
 I(T';K_x')=\frac12\ln\!\left(1+\frac{g^2}{1+b}\right),
 \qquad
 F_{T',K_x'}=\frac{F_Q}{1+b},
\end{align*}
and the mutual information remains unbounded in $g$. Timing jitter behaves
differently. For any fixed $a>0$ and finite $b$, varying $\chi$ at fixed
$\sigma_t$, $\sigma_x$, $\tau$, and $\kappa$ gives
\begin{align}
 \lim_{|g|\to\infty}I(T';K_x')
 &=\frac12\ln\!\left(1+\frac{1}{a}\right)
 =\frac12\ln\!\left(1+\frac{\sigma_t^2}{\tau^2}\right),
 \label{eq:saturation}
\end{align}
independently of $b$. The recorded cross-domain information saturates at
the mutual information $I(T;T')$ between the arrival time and its recorded
value, and a larger coupling cannot raise it further. For $a>0$ and large
$|g|$,
\begin{align}
 \frac{F_{T',K_x'}}{F_Q}&\approx\frac{1+2a}{ag^2},
 \nonumber\\
 \chi^2F_{T',K_x'}&\to2+\frac{\sigma_t^2}{\tau^2},
 \label{eq:jitterFI}
\end{align}
so the relative precision of $\chi$ per recorded outcome becomes
independent of $\chi$.

\subsection{Finite-sample estimation}

The precision for a finite number of samples follows from the same
Gaussian model. For $M$ independent and identically distributed noiseless
outcomes $(T_i,K_{x,i})$ with known zero means, the maximum-likelihood
estimator of the phase-coupling coefficient is the least-squares
regression coefficient
\begin{align*}
 \widehat\chi
 =\frac{\sum_{i=1}^{M}T_iK_{x,i}}{\sum_{i=1}^{M}T_i^2},
\end{align*}
which does not require knowledge of $\sigma_t$ and $\sigma_x$. To leading
order in $1/M$, its variance attains the Cram\'er-Rao bound,
\begin{align}
 \operatorname{Var}(\widehat\chi)
 &\approx\frac{1}{MF_Q}.
 \label{eq:finiteshot}
\end{align}
If $\sigma_t$ and $\sigma_x$ are known, then
$\widehat g=2\sigma_t\sigma_x\widehat\chi$ has
$\operatorname{Var}(\widehat g)\approx1/M$, and first-order error
propagation through Eq.~\eqref{eq:MI} gives the approximate 95\%
confidence interval
\begin{align}
 \widehat I\ \pm\ z_{0.975}\,
 \frac{|\widehat g|}{1+\widehat g^2}\,M^{-1/2},
 \label{eq:intervalknown}
\end{align}
where $\widehat I=\tfrac12\ln(1+\widehat g^{\,2})$ and $z_{0.975}=1.96$ is
the standard-normal quantile. If $\sigma_t$ and $\sigma_x$ are unknown,
the mutual information can be estimated from the sample correlation
coefficient $\widehat r$ as $\widehat I=-\tfrac12\ln(1-\widehat r^{\,2})$,
with $\widehat g\coloneqq\widehat r/\sqrt{1-\widehat r^{\,2}}$. Using
\begin{equation*}
 \operatorname{Var}(\widehat r)\approx\frac{(1-r_{T,K_x}^2)^2}{M}
\end{equation*}
for a bivariate Gaussian sample, the confidence interval becomes
\begin{align}
 \widehat I\ \pm\ z_{0.975}\,
 \frac{|\widehat g|}{\sqrt{1+\widehat g^2}}\,M^{-1/2},
 \label{eq:intervalunknown}
\end{align}
which is wider by the factor $\sqrt{1+g^2}$. For $g=2$ and $M=10^4$, the
half-widths are $7.8\times10^{-3}$ and $1.8\times10^{-2}$~nats,
respectively. These first-order intervals apply for large $M$ and for a
coupling that is separated from zero on its sampling-error scale. At or
near $g=0$ the first derivative of $I$ vanishes and symmetric intervals
are unreliable; a confidence interval for $g$ (or $r$) should instead be
mapped through $I=\tfrac12\ln(1+g^2)$ [or $I=-\tfrac12\ln(1-r^2)$], with
zero as the lower endpoint when the parameter interval contains zero.

\subsection{Estimation under calibrated jitter}

With timing jitter, the regression coefficient of $K_x'$ on $T'$ is biased
by the factor $1/(1+a)$, because
\begin{equation*}
 \operatorname{Cov}(T',K_x')=\chi\sigma_t^2,
 \qquad
 \operatorname{Var}(T')=\sigma_t^2+\tau^2,
\end{equation*}
and it is no longer the maximum-likelihood estimator. If the jitter
variance $\tau^2$ is calibrated independently, for instance from the
instrument response function, the moment estimator
\begin{align}
 \widehat\chi_\tau
 =\frac{\sum_{i=1}^{M}T_i'K_{x,i}'}{\sum_{i=1}^{M}T_i'^{\,2}-M\tau^2}
 \label{eq:chitau}
\end{align}
is consistent and requires no knowledge of $\sigma_t$, $\sigma_x$, or
$\kappa$. Its asymptotic variance exceeds $1/(MF_{T',K_x'})$ in general,
because it ignores the dependence of the residual variance on $\chi$,
which carries the second term of Eq.~\eqref{eq:noisyFI}. To leading
order in $1/M$,
\begin{align*}
 \operatorname{Var}(\widehat\chi_\tau)
 &\approx\frac{\Lambda+2a^2g^2}{MF_Q}
 \\
 &=\frac{1}{MF_{T',K_x'}}\left(1+\frac{2a^2g^2}{\Lambda}\right)^{2},
\end{align*}
and since $\Lambda\ge ag^2$, the excess factor lies between $1$ and
$(1+2a)^2$. Its deviation from unity is of order $a^2g^2$ for $ag^2\ll1$,
and it approaches $(1+2a)^2$ for $ag^2\gg1$. The maximum-likelihood
estimator of the bivariate Gaussian model with known $\tau$ and $\kappa$
attains Eq.~\eqref{eq:noisyFI} asymptotically.

\section{Discussion}
\label{sec:discussion}

\subsection{Design rule}

\subsubsection{Ideal detection}

Equation~\eqref{eq:resourcebound} and Proposition~\ref{prop:optimal} give a
design rule within the canonical model. A nonnegative target cross-domain
information $I_\star$ can be reached with ideal detection by some state if
and only if
\begin{align}
 \beta_T\beta_\nu&\ge\frac{e^{I_\star}}{2},
 &
 \beta_X\beta_K&\ge\frac{e^{I_\star}}{2},
 \label{eq:feasibility}
\end{align}
and, with ideal detection, a sheared Gaussian wave packet reaches it with
\begin{align}
 g_\star&=\sqrt{e^{2I_\star}-1},
 &
 |\chi|&=\frac{g_\star}{2\sigma_t\sigma_x}.
 \label{eq:designshear}
\end{align}
One bit corresponds to $I_\star=\ln2$, which gives $g_\star=\sqrt3$ and
requires each budget product to be at least unity.

\subsubsection{Detector-limited design}

With nonzero detector noise, the coupling $g_\star$ of
Eq.~\eqref{eq:designshear} delivers less than $I_\star$. At chosen input
widths and detector noise, solving Eq.~\eqref{eq:noisyMI} for $g$ gives the
required coupling
\begin{align}
 g_\star^2
 &=\frac{(1+a)(1+b)\bigl(e^{2I_\star}-1\bigr)}
 {1-a\bigl(e^{2I_\star}-1\bigr)},
 \label{eq:noisydesign}
\end{align}
which exists only if $a(e^{2I_\star}-1)<1$, that is, only if $I_\star$ lies
below the saturation level of Eq.~\eqref{eq:saturation}. By
Proposition~\ref{prop:optimal}, the budgets further require
$g_\star^2\le4P_\star^2-1$, and for the chosen widths the individual
conditions
\begin{align*}
 \sigma_t&\le\beta_T, & \sigma_x&\le\beta_X,\\
 \frac{\sqrt{1+g_\star^2}}{2\sigma_t}&\le\beta_\nu,
 &\frac{\sqrt{1+g_\star^2}}{2\sigma_x}&\le\beta_K
\end{align*}
must hold. As an illustration, for one bit with $a=0.04$ and $b=0.1$,
Eq.~\eqref{eq:noisydesign} gives $g_\star=1.97$ instead of $\sqrt3$. The
effective temporal resolution ratio of Eq.~\eqref{eq:slots} is therefore
limited by the smaller of the root-mean-square time-bandwidth and
space-bandwidth products, and for recorded outcomes timing jitter imposes
the additional ceiling of Eq.~\eqref{eq:saturation}.

\subsection{Classical fields and single photons}

The probability density function in Eq.~\eqref{eq:joint} also arises from
a classical coherent field with the same nonseparable spatiotemporal
envelope, with $p(t,k_x)$ read as a normalized intensity distribution.
Lemma~\ref{lem:exclusion} and Propositions~\ref{prop:optimal} and
\ref{prop:noisy} are statements about such normalized distributions and
their Fourier transforms, so they apply to classical wave packets as well.
The non-separability of $\psi_\chi$ between its temporal and spatial
degrees of freedom has the same mathematical form as the non-separability
of classical beams \cite{Spreeuw1998,QianEberly2011,Kondakci2019}; it
involves a single wave packet and is not a nonlocal resource.

When $\psi_\chi$ is the wave function of a single photon, the outcomes are
individual detection events, the R\'enyi-2 entropy of the reduced state is
a monotone of the entanglement between two degrees of freedom of that
photon, and $F_Q$ is the quantum Fisher information per photon. Because
the measured distribution $p(t,k_x)$ is common to both descriptions, a
quantum implementation requires an independent test of the photon
statistics. For a state independently established to be globally pure,
Gaussian in the canonical variables $(T,\nu,X,K_x)$, and generated by the
stated shear, one measured correlation coefficient determines the
R\'enyi-2 entropy and the Schmidt number of the reduced state,
\begin{align}
 S_2(\hat\rho_T)=\ln K
 &=-\frac12\ln\!\left(1-r_{T,K_x}^2\right).
 \label{eq:S2fromr}
\end{align}
This inference relies on the model and does not replace a reconstruction
of the covariance matrix for a general state.

\subsection{Relevance to pulse characterization and sensing}
\label{sec:applications}

\subsubsection{Bounds on spatial side information from second moments}

Because every spatial observable commutes with $\hat T$, a local spatial
measurement preserves the unconditioned distribution of $T$, while its
outcomes may be correlated with $T$. Lemma~\ref{lem:exclusion} bounds the
information that a transverse-wave-number measurement obtains about the
arrival time by $\ln(2\Delta X\Delta K_x)$. For an arbitrary measurement
$E$ on the spatial degree of freedom, the Holevo bound \cite{Holevo1973}
and the maximum-entropy property of Gaussian states at fixed covariance
matrix \cite{wolf2006extremality} give
\begin{align}
 I(T;E)\le S_{\mathrm{vN}}(\hat\rho_X)\le S_{\mathrm G}\bigl(\sqrt{\det V_X}\bigr),
 \label{eq:spatialbound}
\end{align}
where $S_{\mathrm G}(\zeta)$ is the right-hand side of
Eq.~\eqref{eq:vonNeumann}, the entropy of a single-mode Gaussian state with
symplectic eigenvalue $\zeta$. The quantity $\sqrt{\det V_X}$ is invariant
under free paraxial propagation and satisfies
\begin{equation}
 \sqrt{\det V_X}=\frac{\mathcal M_x^2}{2},
 \label{eq:Msquared}
\end{equation}
where $\mathcal M_x^2$ is the second-moment beam-quality factor of the
time-integrated beam. At the beam waist, where $C_{XK_x}=0$, the product
$\Delta X\Delta K_x$ equals $\mathcal M_x^2/2$, so
Lemma~\ref{lem:exclusion} gives $I(T;K_x)\le\ln\mathcal M_x^2$ there. In
other planes $\Delta X\Delta K_x$ is larger, so Lemma~\ref{lem:exclusion}
evaluated there gives a weaker bound. Within the paraxial model, free
propagation acts on the spatial factor as a quadratic phase in $k_x$ and
leaves the joint distribution of $T$ and $K_x$ unchanged, up to the small
$k_x$-dependent delay estimated in Appendix~\ref{app:optical}, so the waist
value applies in every plane; Eq.~\eqref{eq:spatialbound} is plane
independent as well. For a beam with $\mathcal M_x^2=1.2$, any spatial
measurement therefore obtains at most $0.34$~nats about $T$, and a
wave-number measurement at most $\ln1.2=0.18$~nats. Both bounds follow
from second moments obtainable from beam-profile measurements in several
propagation planes, without tomography of the joint state. A related
second-moment quality factor has been used to describe the propagation of
spatial entanglement of photon pairs \cite{Reichert2017}. In time-bin and
temporal-mode encodings \cite{Brecht2015,Raymer2020}, bounds of this type
may be one input to an assessment of spatial side channels
\cite{Rau2014}; such an assessment requires an explicit protocol and
security model, which we do not consider here.

\subsubsection{Time-to-wave-number mapping}

A detector that maps arrival time onto transverse wave number and records
only the latter obtains $I(T;K_x)$ nats about $T$. By
Proposition~\ref{prop:optimal} and Eq.~\eqref{eq:slots}, the Gaussian
temporal resolution ratio is at most $2P_\star$. When the reference time
is recorded with jitter, the mutual information between the two recorded
outcomes is additionally bounded by Eq.~\eqref{eq:saturation}.

\subsubsection{Estimation of spatiotemporal couplings}

Pulse-front tilt and wave-front rotation are characterized routinely in
ultrafast optics \cite{Akturk2005,Gabolde2006,Guang2016}, and the strength
of a continuous-variable controlled-$Z$ interaction sets the edge weights
of cluster states \cite{menicucci2006universal,gu2009quantum}.
Equation~\eqref{eq:classicalFI} shows that recording $(T,K_x)$ attains the
quantum Fisher information for $\chi$ at fixed input widths, which allows
asymptotically efficient estimation with the separable Gaussian probe.
This uses a fixed compatible measurement and complements mode-sorting
approaches to superresolution and time-frequency estimation
\cite{Tsang2016,Donohue2018}. Equation~\eqref{eq:noisyFI} gives the loss
of precision due to timing jitter, Eq.~\eqref{eq:chitau} gives a
consistent estimator under calibrated jitter, and
Eq.~\eqref{eq:jitterFI} sets the asymptotic Cram\'er-Rao limit on the
relative precision per detected photon in this detector model. When the
lens realization is used to estimate an optical tilt parameter, any
accompanying change in the widths must also be included in the
likelihood. At the same centered temporal and spatial variances, a larger
pure-probe $F_Q$ requires correlations between the two degrees of freedom,
as studied in quantum metrology and sensing
\cite{paris2009quantum,degen2017quantum,Descamps2023,LeeNori2023}.

\subsection{Limitations}

The present analysis has several limitations. First, the model on
the whole real line is an approximation; Appendix~\ref{app:model}
quantifies its accuracy for optical pulses. Second, the optimality
statements refer to measurements within the canonical model, in which $T$
is the arrival time at a fixed plane. Third, the detector model of
Sec.~\ref{sec:detector} covers only additive Gaussian noise in the
recorded time and wave number. Other imperfections act differently.
Mode-independent loss reduces the detection rate but leaves the state
conditioned on detection unchanged. A finite aperture with a smooth
transmission modifies the wave packet and can give a pure non-Gaussian
state conditioned on transmission. Dephasing and unresolved timing drift
generally produce mixtures, and multiphoton components lie outside the
single-excitation model. For states in the canonical model that retain the
finite-moment and entropy assumptions, Eqs.~\eqref{eq:generalbound} and
\eqref{eq:spatialbound} remain valid and $I(T;K_x)$ remains a measurable
correlation measure, whereas the equalities in Eqs.~\eqref{eq:identity},
\eqref{eq:schmidt}, \eqref{eq:renyisplit}, and \eqref{eq:S2fromr} need not
hold. Finally, the analysis is carried out at the waveform level; it does
not specify a modulation format, a channel code, a communication rate, or
models of channel noise and synchronization.

\section{Conclusions}
\label{sec:conclusion}

In this work, we have studied the information shared by compatible
observables from different canonical pairs of a wave packet with temporal
and transverse spatial degrees of freedom. By combining entropic
uncertainty relations with the covariance description of
continuous-variable systems, we have shown that, for any state with finite
second moments and differential entropies, the mutual information between
the outcomes of $T$ and $K_x$ is bounded above by the logarithm of twice
the smaller uncertainty product. A Gaussian wave packet with the bilinear
phase $\exp(i\chi\hat T\hat X)$ saturates this bound for both canonical
pairs, with $I(T;K_x)=\tfrac12\ln(1+g^2)$, and maximizes the cross-domain
information over all admissible states under four second-moment budgets.
This attainability yields a necessary and sufficient condition for
reaching a target information within the ideal canonical model.

By analyzing the sheared Gaussian state, we have further shown that local
symplectic transformations bring its covariance matrix to the standard
form of a two-mode squeezed vacuum state and map the physical pair
$(T,K_x)$ to its correlated quadratures. The fixed measurement of
$(T,K_x)$ is the optimal local Gaussian measurement, yields the R\'enyi-2
entropy of the reduced state and the logarithm of the Schmidt number, and
attains the quantum Fisher information for $\chi$ at fixed input widths.
Moreover, with finite detector resolution, the recorded mutual information
is given by Eq.~\eqref{eq:noisyMI}; at fixed widths and with additive
detector noise, it approaches a level set by the ratio of timing jitter to
pulse duration. A classical field with the same envelope produces the same
outcome statistics, whereas a single-photon implementation admits an
interpretation in terms of entanglement between degrees of freedom, once
the photon statistics and the assumptions on the state have been
established independently.

Our work provides a second-moment budget for waveforms that couple
temporal and spatial degrees of freedom, tomography-free bounds on the
information that spatial measurements obtain about the arrival time, and
a fixed measurement that estimates spatiotemporal couplings at the quantum
limit for separable probes. The covariance description used here
treats time and frequency shifts, chirps, dispersion, propagation, lenses,
and the bilinear shear as displacements or symplectic maps, so that the
same budget can be evaluated for other waveform operations. Future work
could address correlated probe states, time-limited and band-limited
signals, and protocol-level communication objectives. In future quantum
communication networks, these budgets could guide the allocation of
time-bandwidth and space-bandwidth resources in spatiotemporally
multiplexed links and help bound spatial side channels in temporal-mode
encodings.

\begin{acknowledgments}
This work was supported by the European Union under the MULTIPLY-6G project (Grant No.~101293106) and the QUEST-6G project (Grant No.~101292676). G.~F. was also supported by the Swedish Foundation for Strategic Research (SSF) through the SAICOM project (Grant No.~FUS21-0004).
F.~N. is supported in part by the Japan Science and Technology Agency (JST) [via the CREST Quantum Frontiers program Grant No.~JPMJCR24I2, the Quantum Leap Flagship Program (Q-LEAP), the Moonshot R\&D Grant No.~JPMJMS256E, and the ASPIRE program (Grant No.~JPMJAP2513)].
M.~R.~L. carried out this work at KTH while on sabbatical leave from IBM Research and is a visiting scientist at the RIKEN Center for Quantum Computing.
\end{acknowledgments}

\section*{Use of AI Tools}

The authors acknowledge the support of Claude Opus $5.5$ and OpenAI Astra $6$ in drafting and editing the text, connecting the results to the literature, and writing the Python scripts. The research idea, lemma and propositions, formulation, results, and conclusions are the authors' own. The authors verified all derivations, numerical values, code, and references, and take full responsibility for the content. No figure was generated or modified by an AI image tool.

\section*{Data and Code Availability}

The data and implementation aspects are described in the text, and are available from the corresponding author upon reasonable request.

\appendix

\section{Model Assumptions for Optical Pulses}
\label{app:model}

\subsection{Operators on the whole real line}

Equation~\eqref{eq:envelope} defines the model on $\mathcal H$. The
operators $\hat T$, $\hat X$, $\hat\nu=-i\partial_t$, and
$\hat K_x=-i\partial_x$ are self-adjoint on $L^2(\mathbb R)$, and their
commutators hold on the Schwartz space of smooth, rapidly decreasing
functions, which contains the Gaussian states used in this work. A
physical optical frequency is positive, so the physical range of $\nu$ is a
half-line, whereas the model takes $\nu$ on the whole real line. The
whole-line model is therefore an approximation, and $\hat T$ acts on the time coordinate of the envelope; we
do not identify it with an operator conjugate to the Hamiltonian of the
field. Its outcome is the arrival time at a fixed detection plane, whose
probability density follows from $|\psi(t,x)|^2$; this is the
photodetection probability density of an ideal broadband detector in the
narrowband limit stated below \cite{Glauber1963}.

\subsection{Positive frequency and transverse propagation}

For one transverse coordinate and a fixed polarization, a single-photon
spatiotemporal wave packet is
\begin{align}
 |\Psi\rangle
 ={}&
 \int_{0}^{\infty}d\omega
 \int_{-k(\omega)}^{k(\omega)}dk_x\,
 \Phi(\omega,k_x)
 \hat a^{\dagger}(\omega,k_x)|0\rangle ,
 \label{eq:opticalstate}
\end{align}
where $|0\rangle$ is the vacuum of the field, $\hat a(\omega,k_x)$ is the
annihilation operator of the plane-wave mode $(\omega,k_x)$ with
\begin{equation*}
 [\hat a(\omega,k_x),\hat a^\dagger(\omega',k_x')]
 =\delta(\omega-\omega')\delta(k_x-k_x'),
\end{equation*}
$\delta(\cdot)$ is the Dirac delta, and $\Phi$ is normalized over the
stated domain. The limits $|k_x|\le k(\omega)=\omega/c$, with $c$ the speed
of light in vacuum, restrict the wave packet to propagating plane waves.
With the convention of Sec.~\ref{sec:canonical_model},
$\nu=\omega_c-\omega$, so the physical range is $\nu<\omega_c$.

Extending $\nu$ and $k_x$ to the whole real line adds negative frequencies
and evanescent components that the physical wave packet does not contain.
This is a good approximation when $\Delta\nu\ll\omega_c$ and
$\Delta K_x\ll k_0$, where $k_0=\omega_c/c$, since the weight that the
Gaussian model assigns to the excluded regions is then exponentially
small. The leading errors are of first order in $\Delta\nu/\omega_c$,
because a detector resolves the propagation angle $\theta$, and the
paraxial relation
\begin{equation*}
 k_x\approx k(\omega)\theta=k_0\left(1-\frac{\nu}{\omega_c}\right)\theta
\end{equation*}
mixes $K_x$ with $\nu$. The paraxial approximation adds corrections of
order $(\Delta K_x/k_0)^2$. In the same limit, the photodetection
probability density of an ideal broadband detector at the detection plane
reduces to $|\psi(t,x)|^2$ up to corrections of order
$\Delta\nu/\omega_c$, which justifies identifying $T$ with the arrival
time.

\section{Optical Realization of the Sheared State}
\label{app:optical}

\subsection{Pulse-front tilt and a lens}

Consider, at an input plane, a pulse with pulse-front tilt,
\begin{align}
 \psi_{\mathrm{in}}(t,x)
 \propto
 \exp\!\left[-\frac{x^2}{4w^2}-\frac{(t+\alpha x)^2}{4\tau_0^2}\right],
 \label{eq:pft}
\end{align}
where $w$ is the root-mean-square beam width, $\tau_0$ is the local
root-mean-square duration, and $\alpha$ is the pulse-front-tilt slope in
seconds per meter. Angular dispersion from a grating or a prism produces
such a field \cite{Akturk2005}. The arrival time and the position at the
input plane obey
\begin{equation*}
 T=-\alpha X+N_t,
 \qquad
 I(T;X)=\tfrac12\ln\!\left(1+\frac{\alpha^2w^2}{\tau_0^2}\right),
\end{equation*}
where $N_t\sim\mathcal N(0,\tau_0^2)$ is independent of $X$.

An ideal lens of focal length $f_L$, with the input and output at its
front and back focal planes, implements the spatial Fourier transform
\begin{equation*}
 \psi_{\mathrm F}(t,x_{\mathrm F})\propto\int dx\,\psi_{\mathrm{in}}(t,x)
 \exp\!\left(-\frac{ik_0xx_{\mathrm F}}{f_L}\right)
\end{equation*}
in the paraxial and narrowband approximations, up to normalization and a
constant phase. The Gaussian integral gives $\psi_{\mathrm F}=\psi_\chi$ of
Eq.~\eqref{eq:state} in the focal-plane coordinate $x_{\mathrm F}$, with
\begin{align}
 g&=\frac{\alpha w}{\tau_0},
 &
 \sigma_t&=\tau_0\sqrt{1+g^2},
 \nonumber\\
 \sigma_x&=\frac{f_L\sqrt{1+g^2}}{2k_0w},
 &
 \chi&=\frac{\alpha k_0w^2}{\tau_0^2f_L(1+g^2)}.
 \label{eq:lensmap}
\end{align}
The focal-plane transverse wave number is $K_x=-(k_0/f_L)X_{\mathrm{in}}$,
where $X_{\mathrm{in}}$ is the position at the input plane. The lens
therefore converts the correlation of $(T,X)$ at the input plane into the
correlation of $(T,K_x)$ at the focal plane, with
$I(T;K_x)_{\mathrm F}=I(T;X)_{\mathrm{in}}$. In the language of
Lemma~\ref{lem:exclusion}, the lens transfers the information from
$I(T;X)$ to $I(T;K_x)$, and pulse-front tilt is converted into wave-front rotation at
the focus \cite{Akturk2005,Vincenti2012,Guang2016}.

The joint measurement of $(T,K_x)$ at the focal plane is equivalent to
recording the arrival time and the transverse position in an image plane of
the input plane. Free propagation over a distance $z$ adds a
$k_x$-dependent group delay of order $z(\Delta K_x/k_0)^2/(2c)$, which is
negligible for the parameters below.

\subsection{Parameter set with a picosecond detector}

Table~\ref{tab:optical} lists one parameter set at a wavelength of
1550~nm with $g=2$. A pulse with $\sigma_t=50$~ps, $\tau_0=22.4$~ps, beam
width $w=10$~mm, and tilt slope $\alpha=4.47$~ps/mm (a pulse-front tilt
angle of $53^\circ$) is focused by a lens with $f_L=0.5$~m. The narrowband
and paraxial parameters are $\Delta\nu/\omega_c=1.8\times10^{-5}$ and
$\Delta K_x/k_0=0.02$, so the approximations of Appendix~\ref{app:model}
hold. The group-delay spread over $z=f_L$ is $0.33$~ps, which is negligible
compared with $\sigma_t$.

A position resolution $\epsilon_x$ (standard deviation), referred to the
input plane, gives
\begin{equation*}
 b=\frac{\epsilon_x^2(1+g^2)}{w^2}.
\end{equation*}
With a timing jitter of $\tau=10$~ps and $\epsilon_x=0.5$~mm, the recorded
information is $0.73$~nats, compared with $0.80$~nats for ideal detection,
and the classical Fisher information is $0.83F_Q$. Superconducting nanowire
detectors reach a timing jitter below 3~ps \cite{korzh2020demonstration},
and single-photon cameras based on the same technology provide position
resolution \cite{oripov2023superconducting}.

\begin{table}[!htbp]
\caption{Optical parameter set for the sheared state at a wavelength of
1550~nm, from Eqs.~\eqref{eq:lensmap}, \eqref{eq:noisyMI},
and \eqref{eq:noisyFI}.}
\label{tab:optical}
\begin{ruledtabular}
\begin{tabular}{lc}
Quantity & Value \\
\hline
$\sigma_t$, $\tau_0$ & 50~ps, 22.4~ps \\
$w$, $\alpha$, $f_L$ & 10~mm, 4.47~ps/mm, 0.5~m \\
$g$, $\chi$ & 2, $1.45\times10^{15}~\mathrm{m^{-1}s^{-1}}$ \\
$\sigma_x$ & 13.8~$\mu$m \\
$\Delta\nu/\omega_c$, $\Delta K_x/k_0$ & $1.8\times10^{-5}$, 0.02 \\
$\tau$, $\epsilon_x$ & 10~ps, 0.5~mm \\
$a$, $b$ & 0.04, 0.0125 \\
$I(T';K_x')$, $\tfrac12\ln(1+1/a)$ & 0.73~nats, 1.63~nats \\
$F_{T',K_x'}/F_Q$ & 0.83
\end{tabular}
\end{ruledtabular}
\end{table}

\subsection{Single photons}

A heralded single photon passed through the same optics has the wave
function $\psi_\chi$, because the optics are linear and act on the
one-photon wave function as on a classical envelope. A measured
second-order intensity correlation $g^{(2)}(0)<1/2$ demonstrates
antibunching of the heralded field and certifies a nonzero single-photon
component; bounding its multiphoton probability
additionally requires the mean photon number or the photon-number
statistics. Space-time wave packets \cite{kondakci2017diffraction} have
been synthesized for heralded single photons \cite{turo2026local}. These
wave packets carry correlations between $\nu$ and $K_x$, which by
Eq.~\eqref{eq:exclT} reduce the budget available to $I(T;K_x)$. The
sheared state instead requires $I(\nu;K_x)=0$ and a correlation between
$\nu$ and $X$.

\section{Derivations for the Gaussian Family}
\label{app:gaussian}

\subsection{Covariance matrices}

In the order $(T,\nu,X,K_x)$, the covariance matrix of the sheared state,
$V_\chi=S_\chi V_0S_\chi^{\top}$, is
\begin{align}
 V_\chi={}&
 \begin{pmatrix}
 \sigma_t^2 & 0 & 0 & \chi\sigma_t^2\\
 0 & \dfrac{1}{4\sigma_t^2}+\chi^2\sigma_x^2
   & \chi\sigma_x^2 & 0\\
 0 & \chi\sigma_x^2 & \sigma_x^2 & 0\\
 \chi\sigma_t^2 & 0 & 0
   & \dfrac{1}{4\sigma_x^2}+\chi^2\sigma_t^2
 \end{pmatrix}.
 \label{eq:appVchi}
\end{align}
The covariance matrices of the reduced states are
\begin{align}
 V_T&=
 \begin{pmatrix}
 \sigma_t^2&0\\
 0&\dfrac{1}{4\sigma_t^2}+\chi^2\sigma_x^2
 \end{pmatrix},
 \nonumber\\
 V_X&=
 \begin{pmatrix}
 \sigma_x^2&0\\
 0&\dfrac{1}{4\sigma_x^2}+\chi^2\sigma_t^2
 \end{pmatrix}.
 \label{eq:appreducedV}
\end{align}
Table~\ref{tab:covariance_summary} collects the determinants and purities.

\begin{table}[!htbp]
\caption{Covariance matrices used in the analysis. The global matrices are
in the order $(T,\nu,X,K_x)$, and each reduced matrix is in the order of
its coordinate and the conjugate variable.}
\label{tab:covariance_summary}
\begin{ruledtabular}
\begin{tabular}{lcc}
Matrix & Determinant & Purity \\
\hline
$V_0$ (separable input) & $1/16$ & $1$ \\
$V_\chi$ (sheared state) & $1/16$ & $1$ \\
$V_T$ (reduced temporal) & $(1+g^2)/4$ & $(1+g^2)^{-1/2}$ \\
$V_X$ (reduced spatial) & $(1+g^2)/4$ & $(1+g^2)^{-1/2}$
\end{tabular}
\end{ruledtabular}
\end{table}

\subsection{Joint density and reduced state}

The joint probability density function of the compatible pair follows from
the unitary Fourier transform with respect to $x$,
\begin{align}
 \check\psi_\chi(t,k_x)
 &\coloneqq
 \frac{1}{\sqrt{2\pi}}
 \int_{\mathbb R}dx\,e^{-ik_xx}\psi_\chi(t,x)
 \nonumber\\
 &=\sqrt{\frac{\sigma_x}{\pi\sigma_t}}
 \exp\!\left[-\frac{t^2}{4\sigma_t^2}
 -\sigma_x^2(k_x-\chi t)^2\right].
 \label{eq:partialFT}
\end{align}
Its squared modulus is Eq.~\eqref{eq:joint}. For fixed $t$, the conditional
spatial state is a pure Gaussian minimum-uncertainty state, and for fixed
$k_x$ the conditional temporal amplitude in Eq.~\eqref{eq:partialFT} is a
real Gaussian function of $t$, which is also a minimum-uncertainty state.
These are the conditions for equality in the entropic uncertainty relation
used in Lemma~\ref{lem:exclusion}.

The partial trace over the spatial degree of freedom gives the reduced
temporal density matrix
\begin{align}
 \rho_T(t,t')
 &\coloneqq
 \int_{\mathbb R}dx\,
 \psi_\chi(t,x)\psi_\chi^*(t',x)
 \nonumber\\
 &=\frac{1}{\sqrt{2\pi}\sigma_t}
 \exp\!\left[-\frac{t^2+t'^2}{4\sigma_t^2}
 -\frac{\chi^2\sigma_x^2}{2}(t-t')^2\right].
 \label{eq:reducedkernel}
\end{align}
With $u=(t+t')/\sqrt2$ and $v=(t-t')/\sqrt2$, the integral for the purity
factorizes,
\begin{align}
 \mu_T
 &=\int_{\mathbb R}dt\int_{\mathbb R}dt'\,
 |\rho_T(t,t')|^2
 \nonumber\\
 &=\frac{1}{2\pi\sigma_t^2}
 \int_{\mathbb R}du\,e^{-u^2/(2\sigma_t^2)}
 \int_{\mathbb R}dv\,
 e^{-[1/(2\sigma_t^2)+2\chi^2\sigma_x^2]v^2}
 \nonumber\\
 &=\frac{1}{\sqrt{1+g^2}}.
 \label{eq:purityintegral}
\end{align}
The same result follows from $\mu_T=1/[2\sqrt{\det V_T}]$
\cite{Weedbrook2012}.

\section{Derivation of Proposition 3}
\label{app:noisy}

Under Eq.~\eqref{eq:noisemodel}, the pair $(T',K_x')$ is a zero-mean
bivariate Gaussian random vector with covariance matrix
\begin{align}
 \Sigma(\chi)
 &=
 \begin{pmatrix}
 \sigma_t^2(1+a) & \chi\sigma_t^2\\[1mm]
 \chi\sigma_t^2 & \dfrac{1+b+g^2}{4\sigma_x^2}
 \end{pmatrix},
 \label{eq:Sigma}
\end{align}
where Eqs.~\eqref{eq:moments}, \eqref{gvar}, and \eqref{eq:ab} have been
used. Its determinant is
\begin{equation*}
 \det\Sigma=\frac{\sigma_t^2\Lambda}{4\sigma_x^2},
 \qquad
 \Lambda=(1+a)(1+b)+ag^2 .
\end{equation*}
The squared correlation coefficient is
\begin{align*}
 r'^{\,2}
 =\frac{g^2}{(1+a)(1+b+g^2)},
\end{align*}
and $I(T';K_x')=-\tfrac12\ln(1-r'^{\,2})$ gives Eq.~\eqref{eq:noisyMI}.

For a zero-mean Gaussian family with covariance matrix $\Sigma(\chi)$, the
Fisher information is
\begin{equation*}
 F=\tfrac12\mathrm{Tr}\!\left[\left(\Sigma^{-1}\partial_\chi\Sigma\right)^2\right].
\end{equation*}
With
\begin{align*}
 \partial_\chi\Sigma
 =\sigma_t^2
 \begin{pmatrix}
 0&1\\
 1&2\chi
 \end{pmatrix},
\end{align*}
a direct evaluation gives
\begin{align*}
 F_{T',K_x'}(\chi)
 &=\frac{4\sigma_t^2\sigma_x^2}{\Lambda}
 +\frac{8\sigma_t^2\sigma_x^2\,a^2g^2}{\Lambda^2},
\end{align*}
which is Eq.~\eqref{eq:noisyFI}. The first term is the Fisher information
carried by the linear regression of $K_x'$ on $T'$, and the second term is
the Fisher information carried by the dependence of the residual variance
on $\chi$.

\section{Proof of Proposition 2}
\label{app:gaussianmeas}

Local symplectic transformations map local Gaussian measurements to local
Gaussian measurements and leave the mutual information of their outcomes
unchanged. It therefore suffices to consider the standard form
$V_{\mathrm{std}}=\bigl(\begin{smallmatrix}A&C\\C^{\top}&B\end{smallmatrix}\bigr)$
of Eq.~\eqref{eq:standardform}, with $A=B=\tfrac{\gamma}{2}\openone_2$ and
$C=\tfrac{g}{2}\mathrm{diag}(1,-1)$.

A Gaussian measurement of the spatial degree of freedom with seed
covariance matrix $W_B$, where $W_B+\tfrac{i}{2}J_1\succeq0$, leaves the
temporal degree of freedom in a Gaussian state whose mean depends linearly
on the outcome and whose covariance matrix
\begin{align}
 \Gamma=A-C(B+W_B)^{-1}C^{\top}
 \label{eq:conditionalcov}
\end{align}
does not depend on the outcome
\cite{giedke2002characterization,Weedbrook2012}. A Gaussian measurement of
the temporal degree of freedom with seed $W_A$ then gives a Gaussian
outcome with covariance matrix $\Gamma+W_A$ conditional on the spatial
outcome and $A+W_A$ unconditionally. Hence
\begin{align}
 I=\frac12\ln\frac{\det(A+W_A)}{\det(\Gamma+W_A)}.
 \label{eq:gaussMI}
\end{align}
Since $W_B\succ0$, $(B+W_B)^{-1}\preceq B^{-1}$, and
\begin{align}
 \Gamma\succeq A-CB^{-1}C^{\top}=\frac{1}{2\gamma}\openone_2.
 \label{eq:schur}
\end{align}
Let $\lambda_1$ and $\lambda_2$ be the eigenvalues of $W_A$, with
$\lambda_1\lambda_2=\det W_A\ge1/4$. Equations~\eqref{eq:gaussMI} and
\eqref{eq:schur} give
\begin{equation*}
 I\le\tfrac12\ln\!\left[\mathcal R(\lambda_1)\mathcal R(\lambda_2)\right],
 \qquad
 \mathcal R(\lambda)\coloneqq\frac{\gamma/2+\lambda}{1/(2\gamma)+\lambda}.
\end{equation*}
The function $\mathcal R$ decreases in $\lambda$, so
$\mathcal R(\lambda_2)\le\mathcal R\bigl(1/(4\lambda_1)\bigr)$, and a
direct expansion gives
\begin{equation*}
 \mathcal R(\lambda)\,\mathcal R\!\left(\frac{1}{4\lambda}\right)=\gamma^2
 \qquad\text{for every }\lambda>0.
\end{equation*}
Therefore $I\le\ln\gamma=\tfrac12\ln(1+g^2)$. A homodyne measurement is
the limit of an infinitely squeezed seed, and the bound holds in that
limit; for homodyne detection of the temporal quadrature along a unit
vector $u$,
\begin{equation*}
 I=\frac12\ln\frac{u^{\top}Au}{u^{\top}\Gamma u}\le\ln\gamma
\end{equation*}
follows directly from Eq.~\eqref{eq:schur}. Equality holds for homodyne
detection of the two correlated position quadratures, which by
Sec.~\ref{sec:shear} is the measurement of $(T,K_x)$. \hfill$\square$

\section{Scope Beyond the Gaussian Family}
\label{app:nongaussian}

\subsection{General separable input}

Let a normalized separable input be $\psi_0(t,x)=f(t)h(x)$ and define
$\widetilde h(k_x)=(2\pi)^{-1/2}\int dx\,e^{-ik_xx}h(x)$. After the
bilinear phase, the joint probability density function of the compatible
pair and the reduced temporal density matrix are
\begin{align}
 p(t,k_x)
 &=|f(t)|^2|\widetilde h(k_x-\chi t)|^2,
 \label{eq:generaljoint}\\
 \rho_T(t,t')
 &=f(t)f^*(t')
 \int_{\mathbb R}dx\,|h(x)|^2e^{i\chi(t-t')x}.
 \label{eq:generalkernel}
\end{align}
The first expression depends on $|\widetilde h|^2$, and the second on the
characteristic function of $|h|^2$. These are different functionals of a
general envelope. Consequently, $I(T;K_x)$ need not equal
$S_2(\hat\rho_T)=\ln K$. The latter equality holds for every pure
bipartite state by the definition of $K$.

\subsection{Continuous-variable example}

Take the Gaussian temporal envelope of Sec.~\ref{sec:shear} and the first
Hermite-Gaussian spatial envelope $h(x)\propto x\exp[-x^2/(4s^2)]$. Then
$(\Delta X)^2=3s^2$, and $K_x=\chi T+Y$ with $Y$ independent of $T$ and
distributed as $|\widetilde h|^2\propto k_x^2\exp(-2s^2k_x^2)$; the
characteristic function of $|h|^2$ is $(1-s^2u^2)\exp(-s^2u^2/2)$. For
$\sigma_t=s=1$ and $\chi=1$, numerical evaluation of
Eqs.~\eqref{eq:generaljoint} and \eqref{eq:generalkernel} gives
$I(T;K_x)=0.701$~nats, $S_2(\hat\rho_T)=\ln K=1.190$~nats, and
$S_{\mathrm{vN}}(\hat\rho_T)=1.411$~nats, whereas
$\ln(2\Delta T\Delta\nu)=1.282$~nats and
$\ln(2\Delta X\Delta K_x)=1.522$~nats. The bound of
Eq.~\eqref{eq:generalbound} holds strictly, and the mutual information
differs from both entropies.

\subsection{Two-level example}

This example lies outside the canonical model, because no operators on a
finite-dimensional space satisfy $[\hat T,\hat\nu]=i\openone$, so
Lemma~\ref{lem:exclusion} does not apply to it; it illustrates only the
distinction between measured correlations and entanglement.

Let $\{|0\rangle_T,|1\rangle_T\}$ and $\{|0\rangle_X,|1\rangle_X\}$ be
orthonormal bases of time bins and position bins with
$\hat T|j\rangle_T=j|j\rangle_T$ and $\hat X|j\rangle_X=j|j\rangle_X$, and
define the discrete Fourier basis states of the spatial degree of freedom,
$|\pm\rangle_X=(|0\rangle_X\pm|1\rangle_X)/\sqrt2$. For $0<\eta<1$, take
\begin{align}
 |\psi_0\rangle
 =\left(\sqrt\eta\,|0\rangle_T+\sqrt{1-\eta}\,|1\rangle_T\right)
 \otimes|+\rangle_X.
\end{align}
The discrete bilinear phase $\hat U_\pi=\exp(i\pi\hat T\hat X)$ produces
\begin{align}
 |\psi_\pi\rangle
 =\sqrt\eta\,|0\rangle_T|+\rangle_X
 +\sqrt{1-\eta}\,|1\rangle_T|-\rangle_X,
 \label{eq:twobinschmidt}
\end{align}
which is a Schmidt decomposition. A measurement of the time bin and of the
spatial degree of freedom in the basis $\{|+\rangle_X,|-\rangle_X\}$, the
discrete counterpart of $K_x$, has perfectly correlated outcomes with
probabilities $\eta$ and $1-\eta$. The Shannon mutual information of these
discrete outcomes and the reduced-state entropies are
\begin{align}
 I(T;K_x)&=S_{\mathrm{vN}}(\hat\rho_T)
 =-\eta\ln\eta-(1-\eta)\ln(1-\eta),
 \label{eq:twobinMI}\\
 S_2(\hat\rho_T)&=-\ln[\eta^2+(1-\eta)^2].
 \label{eq:twobinS2}
\end{align}
Because the measurement is in the Schmidt basis, the mutual information
equals the von Neumann entropy of the reduced state and exceeds $S_2$ for
$\eta\neq1/2$; at $\eta=0.8$ the two values are $0.5004$ and
$0.3857$~nats. In the Gaussian case the ordering is reversed,
$I(T;K_x)=S_2<S_{\mathrm{vN}}$.

The separable mixed state
\begin{align}
 \hat\rho_{\mathrm{sep}}
 ={}&\eta\bigl(|0\rangle\langle0|\bigr)_T
 \otimes\bigl(|+\rangle\langle+|\bigr)_X
 \nonumber\\
 &+(1-\eta)\bigl(|1\rangle\langle1|\bigr)_T
 \otimes\bigl(|-\rangle\langle-|\bigr)_X
 \label{eq:separablecounterexample}
\end{align}
has the same joint statistics for this measurement. The mutual information
is a property of a specified measurement and entanglement is a property of
the full state; their equality in Eq.~\eqref{eq:identity} is a property of
the pure sheared Gaussian state.

\section{Discretized Model, Quantum Circuit, and Numerical Checks}
\label{app:gate}

\begin{figure*}[!ht]
    \centering
    \textbf{(a)}\par\smallskip
    \includegraphics[width=0.46\textwidth]{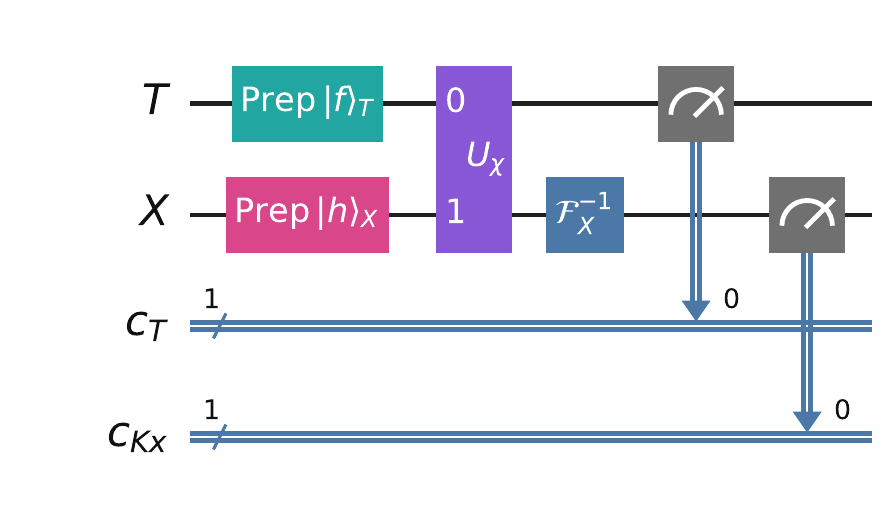}\par\medskip
    \textbf{(b)}\par\smallskip
    \includegraphics[width=0.90\textwidth]{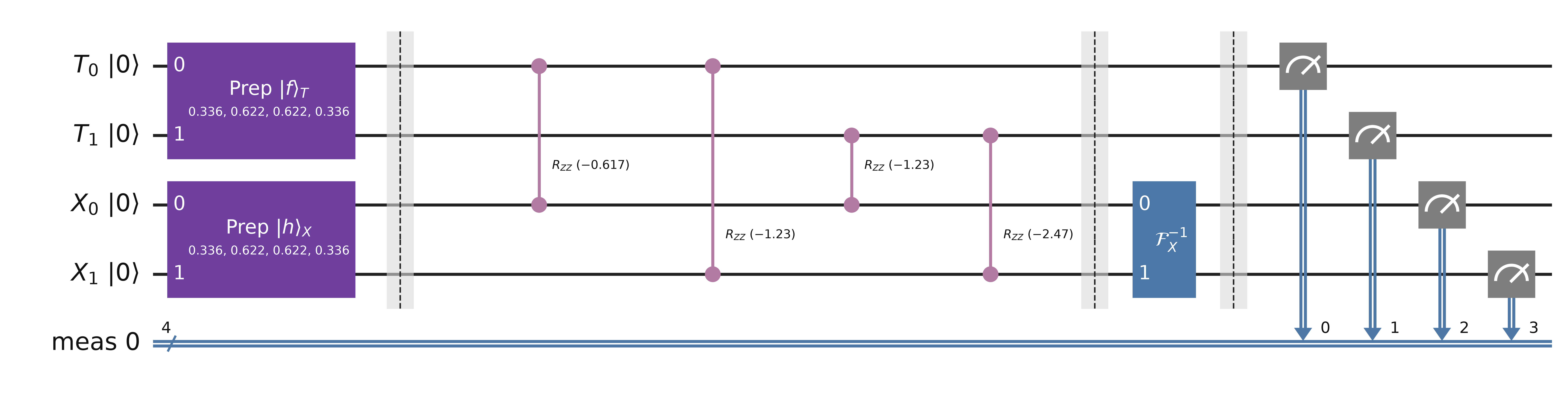}
    \caption{
    Quantum circuit for the temporal-spatial shear in Qiskit.
    (a) Register-level circuit for $n=7$: amplitude encoding of the
    temporal and spatial Gaussian envelopes, the bilinear phase operator
    $U_\chi=\exp(i\chi\hat T\hat X)$, the inverse quantum Fourier transform
    $\mathcal F_X^{-1}$ on the spatial register, after which a
    computational-basis measurement of that register gives the wave number,
    and the measurement of the
    compatible pair. Each wire is a logical register: a quantum wire stands
    for an $n$-qubit register and a classical wire for the corresponding
    $n$-bit register, for 14 qubits in total. Because the drawing
    represents each register by a single wire, the labels 1 and 0 on the
    classical wires are bit labels of that logical wire and do not give
    the register width.
    (b) Four-qubit instance with $n=2$, $d_t=d_x=1$,
    $\sigma_t=\sigma_x=0.9$, and $g=2$, showing the four $R_{ZZ}$ gates with
    rotation angles $\varphi_{qq'}=-\chi d_td_x2^{q+q'-1}$ in radians.
    }
    \label{fig:qiskit_circuits}
\end{figure*}

\subsection{Discretized registers}

To check the continuous-variable formulas on finite registers and to give a
quantum-circuit decomposition of the shear, we discretize the model. The
temporal and spatial coordinates are encoded in separate registers,
\begin{align}
 |f\rangle_T&=\sum_{j=0}^{N-1}f_j|j\rangle_T,
 &
 |h\rangle_X&=\sum_{\ell=0}^{N-1}h_\ell|\ell\rangle_X,
\end{align}
with $n$ qubits per register and $N=2^n$ samples per coordinate. The
centered uniform grids are
\begin{align}
 t_j&=d_t\left[j-\frac{N-1}{2}\right],
 &
 x_\ell&=d_x\left[\ell-\frac{N-1}{2}\right],
 \label{eq:discretizedgrids}
\end{align}
with grid spacings $d_t$ and $d_x$, and the encoded amplitudes are
proportional to $\exp[-t_j^2/(4\sigma_t^2)]$ and
$\exp[-x_\ell^2/(4\sigma_x^2)]$. The reference case uses $n=7$ and the
normalized units
\begin{align}
 d_t&=d_x=1,
 &
 \sigma_t&=\sigma_x=4,
 &
 g&=2,
 &
 \chi&=\frac{1}{16}.
 \label{eq:discretizedparameters}
\end{align}
The wave-number samples are $k_m=2\pi m/(Nd_x)$ with
$m=-N/2,\ldots,N/2-1$, and the discrete Fourier transform is defined by
\begin{align*}
 \langle k_m|x_\ell\rangle
 =N^{-1/2}\exp\!\left(-\frac{2\pi i m\ell}{N}\right),
\end{align*}
with the output reordered so that $m=0$ lies at the center of the grid.
The discrete wave number is defined modulo $2\pi/d_x$, so the shifted
variable $K_x=\chi T+Y$ wraps around when $|\chi t_j|$ approaches the
Nyquist wave number $\pi/d_x$.

\subsection{Gate decomposition of the shear}

In the binary encoding of the coordinates, the centered coordinate
operators are
\begin{align}
 \hat T
 &=-\frac{d_t}{2}
 \sum_{q=0}^{n-1}2^q Z_{T,q},
 &
 \hat X
 &=-\frac{d_x}{2}
 \sum_{q'=0}^{n-1}2^{q'} Z_{X,q'},
 \label{eq:discretizedcoordinates}
\end{align}
where $Z_{T,q}$ and $Z_{X,q'}$ are Pauli-$Z$ operators on qubit $q$ of the
temporal register and qubit $q'$ of the spatial register. With the
two-qubit rotation $R_{ZZ}(\varphi)\coloneqq\exp(-i\varphi Z\otimes Z/2)$,
the bilinear phase operator factorizes exactly, because all the Pauli
products $Z_{T,q}Z_{X,q'}$ commute,
\begin{align}
 U_\chi
 &=
 \prod_{q=0}^{n-1}
 \prod_{q'=0}^{n-1}
 R_{ZZ}(\varphi_{qq'}),
 &
 \varphi_{qq'}&=-\chi d_td_x2^{q+q'-1}.
 \label{eq:discretizedU}
\end{align}
For $n=7$ the shear contains $n^2=49$ two-qubit $R_{ZZ}$ gates, excluding
state preparation and the inverse quantum Fourier transform. An inverse
quantum Fourier transform $\mathcal F_X^{-1}$ on the spatial register maps
$X$ to $K_x$, and a measurement in the computational basis samples the
compatible pair $(T,K_x)$. Figure~\ref{fig:qiskit_circuits} shows the
circuit in Qiskit \cite{JavadiAbhari2024Qiskit}.

\subsection{Numerical checks}

The mutual information is computed from the joint probability mass
function on the $(T,K_x)$ grid, the correlation coefficient from its
second moments, the R\'enyi-2 entropy from the purity of the reduced
temporal density matrix, and the standard deviation $\Delta\nu$ from the
discrete Fourier transform of the reduced temporal state. For the
parameters in Eq.~\eqref{eq:discretizedparameters} we obtain
\begin{align}
 r_{\mathrm{disc}}
 &=0.8944271910,
  \nonumber\\
 I_{\mathrm{disc}}
 &=0.8047189562\ \text{nats},
 \nonumber\\
 S_{2,\mathrm{disc}}
 &=0.8047189562\ \text{nats},
  \nonumber\\
 2\Delta T\Delta\nu
 &=2\Delta X\Delta K_x=2.2360679775.
 \label{eq:discretizedresults}
\end{align}
The corresponding values of the continuous model are $2/\sqrt{5}$,
$\tfrac12\ln5$, and $\sqrt{5}$, and the absolute differences are below
$5\times10^{-15}$. The statevector of the Qiskit circuit agrees with the
direct evaluation of $U_\chi$ and the discrete Fourier transform to within
$2\times10^{-10}$ in every amplitude for all entries of
Table~\ref{tab:convergence}. This agreement checks the discretization and
the analytical formulas; it is not an independent validation of the
physical model.

Table~\ref{tab:convergence} separates the truncation error of the finite
grid from the aliasing error of the discrete Fourier transform. At $d=1$
and $g=2$, increasing $N$ reduces the truncation error until
floating-point precision is reached. At $g=8$ the error is dominated by
aliasing. Increasing $N$ at fixed $d$ does not remove it, because the
Nyquist wave number $\pi/d_x$ is unchanged, whereas halving the grid
spacing reduces it by about six orders of magnitude.

\begin{table}[!tbp]
\caption{Convergence of the discretized model for $\sigma_t=\sigma_x=4$ and
$d_t=d_x\equiv d$. The errors are absolute deviations from
$I=\tfrac12\ln(1+g^2)$ and $r=g/\sqrt{1+g^2}$, and $S_2$ is computed from
the purity of the discrete reduced temporal density matrix.}
\label{tab:convergence}
\footnotesize
\setlength{\tabcolsep}{3pt}
\begin{ruledtabular}
\begin{tabular}{rccccc}
$N$ & $d$ & $g$ & $|\Delta I|$ & $|\Delta r|$ & $|I-S_2|$ \\
\hline
32  & 1.0 & 2 & $1.38\times10^{-3}$  & $7.11\times10^{-4}$  & $1.13\times10^{-3}$ \\
64  & 1.0 & 2 & $9.31\times10^{-14}$ & $3.08\times10^{-14}$ & $8.98\times10^{-14}$ \\
128 & 1.0 & 2 & $<5\times10^{-15}$   & $<5\times10^{-15}$   & $<5\times10^{-15}$ \\
256 & 1.0 & 2 & $<5\times10^{-15}$   & $<5\times10^{-15}$   & $<5\times10^{-15}$ \\
128 & 1.0 & 8 & $5.29\times10^{-3}$  & $3.54\times10^{-2}$  & $5.05\times10^{-3}$ \\
128 & 0.5 & 8 & $1.38\times10^{-9}$  & $3.60\times10^{-8}$  & $1.38\times10^{-9}$ \\
256 & 0.5 & 8 & $1.38\times10^{-9}$  & $3.52\times10^{-8}$  & $1.38\times10^{-9}$
\end{tabular}
\end{ruledtabular}
\end{table}

The statevector simulation uses exact probabilities. The statistical
uncertainty for a finite number of measurement shots is described by
Eqs.~\eqref{eq:finiteshot}, \eqref{eq:intervalknown}, and
\eqref{eq:intervalunknown}. Device noise and execution on quantum hardware
are outside the present model.

\bibliography{reference1}

@article{Gabor1946,
  title   = {{Theory of communication. Part 1: The analysis of information}},
  author  = {Gabor, Dennis},
  journal = {J. Inst. Electr. Eng. Part III: Radio Commun. Eng.},
  volume  = {93},
  number  = {26},
  pages   = {429--441},
  year    = {1946},
  doi     = {10.1049/ji-3-2.1946.0074}
}

@article{Schrodinger1930,
  title   = {{Zum Heisenbergschen Unsch{\"a}rfeprinzip}},
  author  = {Schr{\"o}dinger, Erwin},
  journal = {Sitzungsber. Preuss. Akad. Wiss., Phys.-Math. Kl.},
  volume  = {14},
  pages   = {296--303},
  year    = {1930}
}

@article{Simon1994,
  title   = {{Quantum-noise matrix for multimode systems: U(n) invariance, squeezing, and normal forms}},
  author  = {Simon, R. and Mukunda, N. and Dutta, Biswadeb},
  journal = {Phys. Rev. A},
  volume  = {49},
  number  = {3},
  pages   = {1567--1583},
  year    = {1994},
  doi     = {10.1103/PhysRevA.49.1567}
}

@article{Weedbrook2012,
  title   = {{Gaussian quantum information}},
  author  = {Weedbrook, Christian and Pirandola, Stefano and Garc{\'\i}a-Patr{\'o}n, Ra{\'u}l and Cerf, Nicolas J. and Ralph, Timothy C. and Shapiro, Jeffrey H. and Lloyd, Seth},
  journal = {Rev. Mod. Phys.},
  volume  = {84},
  number  = {2},
  pages   = {621--669},
  year    = {2012},
  doi     = {10.1103/RevModPhys.84.621}
}

@article{hajomer2024continuous,
  title   = {{Continuous-variable quantum passive optical network}},
  author  = {Hajomer, Adnan A. E. and Derkach, Ivan and Filip, Radim and Andersen, Ulrik L. and Usenko, Vladyslav C. and Gehring, Tobias},
  journal = {Light Sci. Appl.},
  volume  = {13},
  pages   = {291},
  year    = {2024},
  doi     = {10.1038/s41377-024-01633-9}
}

@article{schlake2025pulse,
  title   = {{Pulse shape optimization against Doppler shifts and delays in optical quantum communication}},
  author  = {Schlake, Emanuel and Barzel, Roy and R{\"a}tzel, Dennis and L{\"a}mmerzahl, Claus},
  journal = {EPJ Quantum Technol.},
  volume  = {12},
  pages   = {24},
  year    = {2025},
  doi     = {10.1140/epjqt/s40507-025-00321-w}
}

@article{Adesso2012,
  title   = {{Measuring Gaussian quantum information and correlations using the R{\'e}nyi entropy of order 2}},
  author  = {Adesso, Gerardo and Girolami, Davide and Serafini, Alessio},
  journal = {Phys. Rev. Lett.},
  volume  = {109},
  number  = {19},
  pages   = {190502},
  year    = {2012},
  doi     = {10.1103/PhysRevLett.109.190502}
}

@article{Brecht2015,
  title   = {{Photon temporal modes: A complete framework for quantum information science}},
  author  = {Brecht, Benjamin and Reddy, Dileep V. and Silberhorn, Christine and Raymer, Michael G.},
  journal = {Phys. Rev. X},
  volume  = {5},
  number  = {4},
  pages   = {041017},
  year    = {2015},
  doi     = {10.1103/PhysRevX.5.041017}
}

@article{Tsang2016,
  title   = {{Quantum theory of superresolution for two incoherent optical point sources}},
  author  = {Tsang, Mankei and Nair, Ranjith and Lu, Xiao-Ming},
  journal = {Phys. Rev. X},
  volume  = {6},
  number  = {3},
  pages   = {031033},
  year    = {2016},
  doi     = {10.1103/PhysRevX.6.031033}
}

@article{Donohue2018,
  title   = {{Quantum-limited time-frequency estimation through mode-selective photon measurement}},
  author  = {Donohue, John M. and Ansari, Vahid and {\v{R}}eh{\'a}{\v{c}}ek, Jaroslav and Hradil, Zden{\v{e}}k and Stoklasa, Bohumil and Pa{\'u}r, Martin and S{\'a}nchez-Soto, Luis L. and Silberhorn, Christine},
  journal = {Phys. Rev. Lett.},
  volume  = {121},
  number  = {9},
  pages   = {090501},
  year    = {2018},
  doi     = {10.1103/PhysRevLett.121.090501}
}

@article{Akturk2005,
  title   = {{The general theory of first-order spatio-temporal distortions of Gaussian pulses and beams}},
  author  = {Akturk, Selcuk and Gu, Xun and Gabolde, Pablo and Trebino, Rick},
  journal = {Opt. Express},
  volume  = {13},
  number  = {21},
  pages   = {8642--8661},
  year    = {2005},
  doi     = {10.1364/OPEX.13.008642}
}

@article{Lami2023,
  title   = {{Fundamental limitations to key distillation from Gaussian states with Gaussian operations}},
  author  = {Lami, Ludovico and Mi{\v{s}}ta, Jr., Ladislav and Adesso, Gerardo},
  journal = {Phys. Rev. Res.},
  volume  = {5},
  number  = {3},
  pages   = {033153},
  year    = {2023},
  doi     = {10.1103/PhysRevResearch.5.033153}
}

@article{Guang2016,
  title   = {{Measurement of the ultrafast lighthouse effect using a complete spatiotemporal pulse-characterization technique}},
  author  = {Guang, Zhe and Rhodes, Michelle and Trebino, Rick},
  journal = {J. Opt. Soc. Am. B},
  volume  = {33},
  number  = {9},
  pages   = {1955--1962},
  year    = {2016},
  doi     = {10.1364/JOSAB.33.001955}
}

@article{Fabre2022,
  title   = {{Time and frequency as quantum continuous variables}},
  author  = {Fabre, Nicolas and Keller, Arne and Milman, P{\'e}rola},
  journal = {Phys. Rev. A},
  volume  = {105},
  number  = {5},
  pages   = {052429},
  year    = {2022},
  doi     = {10.1103/PhysRevA.105.052429}
}

@article{shendryk2025efficient,
  title   = {{Efficient implementation of quantum signal processing via the adiabatic-impulse model}},
  author  = {Shendryk, D. O. and Ivakhnenko, O. V. and Shevchenko, S. N. and Nori, Franco},
  journal = {Phys. Rev. A},
  volume  = {112},
  number  = {4},
  pages   = {042437},
  year    = {2025},
  doi     = {10.1103/gnr5-583s}
}

@article{Descamps2023,
  title   = {{Quantum metrology using time-frequency as quantum continuous variables: Resources, sub-shot-noise precision and phase space representation}},
  author  = {Descamps, Eloi and Fabre, Nicolas and Keller, Arne and Milman, P{\'e}rola},
  journal = {Phys. Rev. Lett.},
  volume  = {131},
  number  = {3},
  pages   = {030801},
  year    = {2023},
  doi     = {10.1103/PhysRevLett.131.030801}
}

@article{turo2026local,
  title   = {{Local and remote synthesis of single-photon space-time wave packets}},
  author  = {Turo, Bryan L. and Saleh, Bahaa E. A. and Abouraddy, Ayman F.},
  journal = {Nat. Photonics},
  volume  = {20},
  pages   = {539--548},
  year    = {2026},
  doi     = {10.1038/s41566-026-01872-9}
}

@article{Glauber1963,
  title   = {{The quantum theory of optical coherence}},
  author  = {Glauber, Roy J.},
  journal = {Phys. Rev.},
  volume  = {130},
  number  = {6},
  pages   = {2529--2539},
  year    = {1963},
  doi     = {10.1103/PhysRev.130.2529}
}

@article{MugaLeavens2000,
  title   = {{Arrival time in quantum mechanics}},
  author  = {Muga, J. G. and Leavens, C. R.},
  journal = {Phys. Rep.},
  volume  = {338},
  number  = {4},
  pages   = {353--438},
  year    = {2000},
  doi     = {10.1016/S0370-1573(00)00047-8}
}

@article{MistaTatham2016,
  title   = {{Gaussian intrinsic entanglement}},
  author  = {Mi{\v{s}}ta, Jr., Ladislav and Tatham, Richard},
  journal = {Phys. Rev. Lett.},
  volume  = {117},
  number  = {24},
  pages   = {240505},
  year    = {2016},
  doi     = {10.1103/PhysRevLett.117.240505}
}

@article{eldar2002quantum,
  title   = {{Quantum signal processing}},
  author  = {Eldar, Yonina C. and Oppenheim, Alan V.},
  journal = {IEEE Signal Process. Mag.},
  volume  = {19},
  number  = {6},
  pages   = {12--32},
  year    = {2002},
  doi     = {10.1109/MSP.2002.1043298}
}

@article{low2017optimal,
  title   = {{Optimal Hamiltonian simulation by quantum signal processing}},
  author  = {Low, Guang Hao and Chuang, Isaac L.},
  journal = {Phys. Rev. Lett.},
  volume  = {118},
  number  = {1},
  pages   = {010501},
  year    = {2017},
  doi     = {10.1103/PhysRevLett.118.010501}
}

@inproceedings{gilyen2019quantum,
  title     = {{Quantum singular value transformation and beyond: Exponential improvements for quantum matrix arithmetics}},
  author    = {Gily{\'e}n, Andr{\'a}s and Su, Yuan and Low, Guang Hao and Wiebe, Nathan},
  booktitle = {Proceedings of the 51st Annual ACM SIGACT Symposium on Theory of Computing (STOC '19)},
  publisher = {ACM},
  address   = {New York},
  pages     = {193--204},
  year      = {2019},
  doi       = {10.1145/3313276.3316366}
}

@article{braunstein2005quantum,
  title   = {{Quantum information with continuous variables}},
  author  = {Braunstein, Samuel L. and van Loock, Peter},
  journal = {Rev. Mod. Phys.},
  volume  = {77},
  number  = {2},
  pages   = {513--577},
  year    = {2005},
  doi     = {10.1103/RevModPhys.77.513}
}

@article{gu2009quantum,
  title   = {{Quantum computing with continuous-variable clusters}},
  author  = {Gu, Mile and Weedbrook, Christian and Menicucci, Nicolas C. and Ralph, Timothy C. and van Loock, Peter},
  journal = {Phys. Rev. A},
  volume  = {79},
  number  = {6},
  pages   = {062318},
  year    = {2009},
  doi     = {10.1103/PhysRevA.79.062318}
}

@article{adesso2014continuous,
  title   = {{Continuous variable quantum information: Gaussian states and beyond}},
  author  = {Adesso, Gerardo and Ragy, Sammy and Lee, Antony R.},
  journal = {Open Syst. Inf. Dyn.},
  volume  = {21},
  number  = {01n02},
  pages   = {1440001},
  year    = {2014},
  doi     = {10.1142/S1230161214400010}
}

@article{DresselNori2014,
  title   = {{Certainty in Heisenberg's uncertainty principle: Revisiting definitions for estimation errors and disturbance}},
  author  = {Dressel, Justin and Nori, Franco},
  journal = {Phys. Rev. A},
  volume  = {89},
  number  = {2},
  pages   = {022106},
  year    = {2014},
  doi     = {10.1103/PhysRevA.89.022106}
}

@article{LeeNori2023,
  title   = {{Steering-enhanced quantum metrology using superpositions of noisy phase shifts}},
  author  = {Lee, Kuan-Yi and Lin, Jhen-Dong and Miranowicz, Adam and Nori, Franco and Ku, Huan-Yu and Chen, Yueh-Nan},
  journal = {Phys. Rev. Res.},
  volume  = {5},
  number  = {1},
  pages   = {013103},
  year    = {2023},
  doi     = {10.1103/PhysRevResearch.5.013103}
}

@article{BialynickiBirula1975,
  title   = {{Uncertainty relations for information entropy in wave mechanics}},
  author  = {Bia{\l}ynicki-Birula, Iwo and Mycielski, Jerzy},
  journal = {Commun. Math. Phys.},
  volume  = {44},
  number  = {2},
  pages   = {129--132},
  year    = {1975},
  doi     = {10.1007/BF01608825}
}

@article{Hall1995,
  title   = {{Information exclusion principle for complementary observables}},
  author  = {Hall, Michael J. W.},
  journal = {Phys. Rev. Lett.},
  volume  = {74},
  number  = {17},
  pages   = {3307--3311},
  year    = {1995},
  doi     = {10.1103/PhysRevLett.74.3307}
}

@article{Walborn2011,
  title   = {{Revealing hidden Einstein-Podolsky-Rosen nonlocality}},
  author  = {Walborn, S. P. and Salles, A. and Gomes, R. M. and Toscano, F. and {Souto Ribeiro}, P. H.},
  journal = {Phys. Rev. Lett.},
  volume  = {106},
  number  = {13},
  pages   = {130402},
  year    = {2011},
  doi     = {10.1103/PhysRevLett.106.130402}
}

@article{Coles2017,
  title   = {{Entropic uncertainty relations and their applications}},
  author  = {Coles, Patrick J. and Berta, Mario and Tomamichel, Marco and Wehner, Stephanie},
  journal = {Rev. Mod. Phys.},
  volume  = {89},
  number  = {1},
  pages   = {015002},
  year    = {2017},
  doi     = {10.1103/RevModPhys.89.015002}
}

@article{Furrer2014,
  title   = {{Position-momentum uncertainty relations in the presence of quantum memory}},
  author  = {Furrer, Fabian and Berta, Mario and Tomamichel, Marco and Scholz, Volkher B. and Christandl, Matthias},
  journal = {J. Math. Phys.},
  volume  = {55},
  number  = {12},
  pages   = {122205},
  year    = {2014},
  doi     = {10.1063/1.4903989}
}

@article{BraunsteinCaves1994,
  title   = {{Statistical distance and the geometry of quantum states}},
  author  = {Braunstein, Samuel L. and Caves, Carlton M.},
  journal = {Phys. Rev. Lett.},
  volume  = {72},
  number  = {22},
  pages   = {3439--3443},
  year    = {1994},
  doi     = {10.1103/PhysRevLett.72.3439}
}

@article{Fedorov2004,
  title   = {{Packet narrowing and quantum entanglement in photoionization and photodissociation}},
  author  = {Fedorov, M. V. and Efremov, M. A. and Kazakov, A. E. and Chan, K. W. and Law, C. K. and Eberly, J. H.},
  journal = {Phys. Rev. A},
  volume  = {69},
  number  = {5},
  pages   = {052117},
  year    = {2004},
  doi     = {10.1103/PhysRevA.69.052117}
}

@article{LawEberly2004,
  title   = {{Analysis and interpretation of high transverse entanglement in optical parametric down conversion}},
  author  = {Law, C. K. and Eberly, J. H.},
  journal = {Phys. Rev. Lett.},
  volume  = {92},
  number  = {12},
  pages   = {127903},
  year    = {2004},
  doi     = {10.1103/PhysRevLett.92.127903}
}

@article{Spreeuw1998,
  title   = {{A classical analogy of entanglement}},
  author  = {Spreeuw, Robert J. C.},
  journal = {Found. Phys.},
  volume  = {28},
  number  = {3},
  pages   = {361--374},
  year    = {1998},
  doi     = {10.1023/A:1018703709245}
}

@article{QianEberly2011,
  title   = {{Entanglement and classical polarization states}},
  author  = {Qian, Xiao-Feng and Eberly, J. H.},
  journal = {Opt. Lett.},
  volume  = {36},
  number  = {20},
  pages   = {4110--4112},
  year    = {2011},
  doi     = {10.1364/OL.36.004110}
}

@article{Kondakci2019,
  title   = {{Classical entanglement underpins the invariant propagation of space-time wave packets}},
  author  = {Kondakci, H. Esat and Alonso, Miguel A. and Abouraddy, Ayman F.},
  journal = {Opt. Lett.},
  volume  = {44},
  number  = {11},
  pages   = {2645},
  year    = {2019},
  doi     = {10.1364/OL.44.002645}
}

@book{Cover2006,
  title     = {{Elements of Information Theory}},
  author    = {Cover, Thomas M. and Thomas, Joy A.},
  edition   = {2nd},
  publisher = {Wiley-Interscience},
  address   = {Hoboken, NJ},
  year      = {2006}
}

@article{Vincenti2012,
  title   = {{Attosecond lighthouses: How to use spatiotemporally coupled light fields to generate isolated attosecond pulses}},
  author  = {Vincenti, H. and Qu{\'e}r{\'e}, F.},
  journal = {Phys. Rev. Lett.},
  volume  = {108},
  number  = {11},
  pages   = {113904},
  year    = {2012},
  doi     = {10.1103/PhysRevLett.108.113904}
}

@misc{JavadiAbhari2024Qiskit,
  title         = {{Quantum computing with Qiskit}},
  author        = {Javadi-Abhari, Ali and Treinish, Matthew and Krsulich, Kevin and Wood, Christopher J. and Lishman, Jake and Gacon, Julien and Martiel, Simon and Nation, Paul D. and Bishop, Lev S. and Cross, Andrew W. and Johnson, Blake R. and Gambetta, Jay M.},
  year          = {2024},
  eprint        = {2405.08810},
  archivePrefix = {arXiv},
  primaryClass  = {quant-ph}
}

@article{Holevo1973,
  title   = {{Bounds for the quantity of information transmitted by a quantum communication channel}},
  author  = {Holevo, A. S.},
  journal = {Probl. Inf. Transm.},
  volume  = {9},
  number  = {3},
  pages   = {177--183},
  year    = {1973}
}

@article{Gabolde2006,
  title   = {{Single-shot measurement of the full spatio-temporal field of ultrashort pulses with multi-spectral digital holography}},
  author  = {Gabolde, Pablo and Trebino, Rick},
  journal = {Opt. Express},
  volume  = {14},
  number  = {23},
  pages   = {11460--11467},
  year    = {2006}
}

@article{gouzien2018quantum,
  title   = {{Quantum description of timing jitter for single-photon on-off detectors}},
  author  = {Gouzien, {\'E}lie and Fedrici, Bruno and Zavatta, Alessandro and Tanzilli, S{\'e}bastien and D'Auria, Virginia},
  journal = {Phys. Rev. A},
  volume  = {98},
  number  = {1},
  pages   = {013833},
  year    = {2018},
  doi     = {10.1103/PhysRevA.98.013833}
}

@article{Ansari2018,
  title   = {{Tailoring nonlinear processes for quantum optics with pulsed temporal-mode encodings}},
  author  = {Ansari, Vahid and Donohue, John M. and Brecht, Benjamin and Silberhorn, Christine},
  journal = {Optica},
  volume  = {5},
  number  = {5},
  pages   = {534--550},
  year    = {2018},
  doi     = {10.1364/OPTICA.5.000534}
}

@article{Raymer2020,
  title   = {{Temporal modes in quantum optics: Then and now}},
  author  = {Raymer, Michael G. and Walmsley, Ian A.},
  journal = {Phys. Scr.},
  volume  = {95},
  number  = {6},
  pages   = {064002},
  year    = {2020},
  doi     = {10.1088/1402-4896/ab6153}
}

@article{Rau2014,
  title   = {{Spatial mode side channels in free-space QKD implementations}},
  author  = {Rau, Markus and Vogl, Tobias and Corrielli, Giacomo and Vest, Gwenaelle and Fuchs, Lukas and Nauerth, Sebastian and Weinfurter, Harald},
  journal = {IEEE J. Sel. Top. Quantum Electron.},
  volume  = {21},
  number  = {3},
  pages   = {6600905},
  year    = {2015},
  doi     = {10.1109/JSTQE.2014.2372008}
}

@article{Robertson1929,
  title   = {{The uncertainty principle}},
  author  = {Robertson, H. P.},
  journal = {Phys. Rev.},
  volume  = {34},
  number  = {1},
  pages   = {163--164},
  year    = {1929},
  doi     = {10.1103/PhysRev.34.163}
}

@article{coles2014improved,
  title   = {{Improved entropic uncertainty relations and information exclusion relations}},
  author  = {Coles, Patrick J. and Piani, Marco},
  journal = {Phys. Rev. A},
  volume  = {89},
  number  = {2},
  pages   = {022112},
  year    = {2014},
  doi     = {10.1103/PhysRevA.89.022112}
}

@article{paris2009quantum,
  title   = {{Quantum estimation for quantum technology}},
  author  = {Paris, Matteo G. A.},
  journal = {Int. J. Quantum Inf.},
  volume  = {7},
  number  = {supp01},
  pages   = {125--137},
  year    = {2009},
  doi     = {10.1142/S0219749909004839}
}

@article{degen2017quantum,
  title   = {{Quantum sensing}},
  author  = {Degen, C. L. and Reinhard, F. and Cappellaro, P.},
  journal = {Rev. Mod. Phys.},
  volume  = {89},
  number  = {3},
  pages   = {035002},
  year    = {2017},
  doi     = {10.1103/RevModPhys.89.035002}
}

@article{giedke2002characterization,
  title   = {{Characterization of Gaussian operations and distillation of Gaussian states}},
  author  = {Giedke, G{\'e}za and Cirac, J. Ignacio},
  journal = {Phys. Rev. A},
  volume  = {66},
  number  = {3},
  pages   = {032316},
  year    = {2002},
  doi     = {10.1103/PhysRevA.66.032316}
}

@article{wolf2006extremality,
  title   = {{Extremality of Gaussian quantum states}},
  author  = {Wolf, Michael M. and Giedke, Geza and Cirac, J. Ignacio},
  journal = {Phys. Rev. Lett.},
  volume  = {96},
  number  = {8},
  pages   = {080502},
  year    = {2006},
  doi     = {10.1103/PhysRevLett.96.080502}
}

@article{menicucci2006universal,
  title   = {{Universal quantum computation with continuous-variable cluster states}},
  author  = {Menicucci, Nicolas C. and van Loock, Peter and Gu, Mile and Weedbrook, Christian and Ralph, Timothy C. and Nielsen, Michael A.},
  journal = {Phys. Rev. Lett.},
  volume  = {97},
  number  = {11},
  pages   = {110501},
  year    = {2006},
  doi     = {10.1103/PhysRevLett.97.110501}
}

@article{gatti2009x,
  title   = {{X entanglement: The nonfactorable spatiotemporal structure of biphoton correlation}},
  author  = {Gatti, A. and Brambilla, E. and Caspani, L. and Jedrkiewicz, O. and Lugiato, L. A.},
  journal = {Phys. Rev. Lett.},
  volume  = {102},
  number  = {22},
  pages   = {223601},
  year    = {2009},
  doi     = {10.1103/PhysRevLett.102.223601}
}

@article{kondakci2017diffraction,
  title   = {{Diffraction-free space-time light sheets}},
  author  = {Kondakci, H. Esat and Abouraddy, Ayman F.},
  journal = {Nat. Photonics},
  volume  = {11},
  number  = {11},
  pages   = {733--740},
  year    = {2017},
  doi     = {10.1038/s41566-017-0028-9}
}

@article{korzh2020demonstration,
  title   = {{Demonstration of sub-3 ps temporal resolution with a superconducting nanowire single-photon detector}},
  author  = {Korzh, Boris and Zhao, Qing-Yuan and Allmaras, Jason P. and Frasca, Simone and Autry, Travis M. and Bersin, Eric A. and Beyer, Andrew D. and Briggs, Ryan M. and Bumble, Bruce and Colangelo, Marco and others},
  journal = {Nat. Photonics},
  volume  = {14},
  number  = {4},
  pages   = {250--255},
  year    = {2020},
  doi     = {10.1038/s41566-020-0589-x}
}

@article{oripov2023superconducting,
  title   = {{A superconducting nanowire single-photon camera with 400,000 pixels}},
  author  = {Oripov, Bakhrom G. and Rampini, Dana S. and Allmaras, Jason and Shaw, Matthew D. and Nam, Sae Woo and Korzh, Boris and McCaughan, Adam N.},
  journal = {Nature},
  volume  = {622},
  number  = {7984},
  pages   = {730--734},
  year    = {2023},
  doi     = {10.1038/s41586-023-06550-2}
}

@article{mivsta2011measurement,
  title   = {{Measurement-induced disturbances and nonclassical correlations of Gaussian states}},
  author  = {Mi{\v{s}}ta, Jr., Ladislav and Tatham, Richard and Girolami, Davide and Korolkova, Natalia and Adesso, Gerardo},
  journal = {Phys. Rev. A},
  volume  = {83},
  number  = {4},
  pages   = {042325},
  year    = {2011},
  doi     = {10.1103/PhysRevA.83.042325}
}

@article{Tasca2011,
  title   = {{Continuous-variable quantum computation with spatial degrees of freedom of photons}},
  author  = {Tasca, D. S. and Gomes, R. M. and Toscano, F. and {Souto Ribeiro}, P. H. and Walborn, S. P.},
  journal = {Phys. Rev. A},
  volume  = {83},
  number  = {5},
  pages   = {052325},
  year    = {2011},
  doi     = {10.1103/PhysRevA.83.052325}
}

@article{Reichert2017,
  title   = {{Quality of spatial entanglement propagation}},
  author  = {Reichert, Matthew and Sun, Xiaohang and Fleischer, Jason W.},
  journal = {Phys. Rev. A},
  volume  = {95},
  number  = {6},
  pages   = {063836},
  year    = {2017},
  doi     = {10.1103/PhysRevA.95.063836}
}
\end{document}